%% file: CGO_2018_arXiv.tex
\documentclass[
	sigplan,
	10pt,
]{acmart}

\usepackage{booktabs} 
\usepackage{listings}

\lstdefinestyle{mystyle}{
	language=Python,
	columns=[c]fullflexible,
	mathescape=true,
	tabsize=2,
	breaklines=true,
	lineskip=1pt,
	morekeywords={yield},
	escapeinside={(*}{*)}
}
\newcommand{\code}[1]{{\texttt{#1}}}

\usepackage{drawmatrix}

\usepackage{tikz}
\usetikzlibrary{positioning}
\usetikzlibrary{shapes.geometric}
\usetikzlibrary{plotmarks}
\tikzset{tbox/.style={fill=rwthblack10, rounded corners}}
\tikzset{barrow/.style={line width=0.4mm}}

\usepackage{pgfplots}
\usepackage{pgfplotstable}
\pgfplotsset{compat=newest}

\usepackage{balance}

\input{RWTH_colors_v2}

\setcopyright{acmlicensed}
\acmPrice{15.00}
\acmDOI{10.1145/3168804}
\acmYear{2018}
\copyrightyear{2018}
\acmISBN{978-1-4503-5617-6/18/02}
\acmConference[CGO'18]{2018 IEEE/ACM International Symposium on Code Generation and Optimization}{February 24--28, 2018}{Vienna, Austria}

\begin{document}
\title{The Generalized Matrix Chain Algorithm}

\author{Henrik Barthels}
 \affiliation{
   \department{AICES}              
   \institution{RWTH Aachen University}            
   \streetaddress{Schinkelstr. 2}
   \city{Aachen}
   \state{NRW}
   \postcode{52062}
   \country{Germany}
 }
\email{barthels@aices.rwth-aachen.de}          

\author{Marcin Copik}
 \affiliation{
   \department{AICES}             
   \institution{RWTH Aachen University}           
   \streetaddress{Schinkelstr. 2}
   \city{Aachen}
   \state{NRW}
   \postcode{52062}
   \country{Germany}
 }
\email{mcopik@gmail.com}         

\author{Paolo Bientinesi}
 \affiliation{
   \department{AICES}             
   \institution{RWTH Aachen University}           
   \streetaddress{Schinkelstr. 2}
   \city{Aachen}
   \state{NRW}
   \postcode{52062}
   \country{Germany}
 }
\email{pauldj@aices.rwth-aachen.de}         

\renewcommand{\shorttitle}{The Generalized Matrix Chain Algorithm}

\begin{abstract}

In this paper, we present a generalized version of the matrix chain algorithm to generate efficient code for linear algebra problems, a task for which human experts often invest days or even weeks of works.
The standard matrix chain problem consists in finding the parenthesization of a matrix product $M := A_1 A_2 \cdots A_n$ that minimizes the number of scalar operations. In practical applications, however, one frequently encounters more complicated expressions, involving transposition, inversion, and matrix properties. Indeed, the computation of such expressions relies on a set of computational kernels that offer functionality well beyond the simple matrix product. The challenge then shifts from finding an optimal parenthesization to finding an optimal mapping of the input expression to the available kernels. Furthermore, it is often the case that a solution based on the minimization of scalar operations does not result in the optimal solution in terms of execution time. In our experiments, the generated code outperforms other libraries and languages on average by a factor of about 9. The motivation for this work comes from the fact that---despite great advances in the development of compilers---the task of mapping linear algebra problems to optimized kernels is still to be done manually. In order to relieve the user from this complex task, new techniques for the compilation of linear algebra expressions have to be developed.

\end{abstract}

%
%





\maketitle

\section{Introduction}

Although the evolution of languages and compilers in the last 60 years is
nothing short of remarkable, when it comes to linear algebra computations, the
efficiency levels achieved by experts are still unmatched. 
Popular systems such as Matlab \cite{matlabdoc} and Julia
\cite{bezanson2012} allow to directly express matrix expression, which are
however evaluated according to general and simple rules, usually resulting in
slow execution. To achieve optimal performance, it is necessary to manually
map the expressions to optimized routines for basic linear algebra operations,
for example as provided by libraries such as BLAS \cite{lawson1979,
  dongarra1988, dongarra1990} and LAPACK \cite{anderson1999}. To automate this
task, the linear algebra compiler Linnea is developed
\cite{barthels2016src}.

With Linnea, the objective is twofold: On the one hand, the goal is to generate programs that come close to the efficiency achieved by human experts; on the
other hand, the goal is also to achive a generation time that is only a fraction of what
a human would need, and without requiring the user to possess expertise in
linear algebra or high-per\-for\-mance computing.
As input, Linnea takes expressions as described by the grammar shown in
Fig.~\ref{fig:grammar}, in combination with a description of the operands and
their properties, as shown in Fig.~\ref{fig:grammarops}.
In this paper, we are concerned with expressions consisting of products of
matrices and vectors; we also allow each of the operands to carry properties,
and to be transposed and/or inverted. As examples, consider $X := A B^T C$ and  $x := A^{-1} B y$,
where $A, B, C, X$ are matrices, and $x$ and $y$ are vectors; to be computed
efficiently, these expressions have to be mapped onto a set $K$ of
computational kernels\footnote{The terminology is explained in Sec.~\ref{sec:terminology}.} (e.g.: {\tt C:=A*B}, {\tt C:= A$^\text{\tt -1}$*B}, {\tt
  B:= A$^\text{\tt -1}$}, \dots). Furthermore, the mapping has to minimize a user-selected
cost metric (such as number of flops or execution time). The output is then a sequence of kernel calls that computes the original expression.

\begin{figure}[]
\begin{minipage}{\columnwidth}
\begin{align*}
	\text{assignments} &\rightarrow \text{assignment}^{+}\\
	\text{assignment} &\rightarrow \textit{symbol} := \text{expr}\\
	\text{expr} &\rightarrow \textit{symbol} \mid \text{expr} + \text{expr} \mid \text{expr} \cdot \text{expr} \mid\\
		&\qquad  \text{expr}^{-1} \mid \text{expr}^{T} \mid \text{expr}^{-T}
\end{align*}
\end{minipage}
    \caption{Linnea grammar describing the definition of expressions.}
    \label{fig:grammar}
\end{figure}

We refer to this problem as the \emph{Generalized Matrix Chain Problem}
(GMCP); the classic Matrix Chain Problem (MCP) covers the specific instances
of GMCP in which the input expression only consists of products (without
additional operators such as transposition and inversion), $K$ contains only
the kernel {\tt C:=A*B}, and none of the matrices have a special structure or
  properties \cite{cormen1990}. Thus, MCP only consists in finding an optimal parenthesization. The number of floating point operations (FLOPs) is used as a cost metric.\footnote{The cost of a product $AB$, with $A \in
  \mathbb{R}^{n \times k}$ and $B \in \mathbb{R}^{k \times m}$, is $2 m n k$
  FLOPs.} The applicability of the original algorithm, while well suited to
study dynamic programming algorithms, is fairly limited. Having analyzed
dozens of linear algebra algorithms, we observed that long matrix chains occur
only rarely. On the other hand, expressions involving the product of up to about ten matrices, where some of them are transposed and/or inverted, are much more common. These more complicated matrix chains occur as part of linear algebra problems in many different fields.
For example, $L_{22}^{-1} L_{21} L_{11}^{-1} L_{10}$ is part of a blocked algorithm for the inversion of a triangular matrix \cite{bientinesi2008}. The chains $X^b_i S_i (Y^b_i)^T R_i^{-1}$ is encountered in the ensemble Kalman filter \cite{rao2015:siam}. In an algorithm to reduce matrices to tridiagonal form, $\tau_v \tau_v v v^T A u u^T$ appears \cite{choi1995}. Blocked algorithms for the simultaneous solution of two linear systems, $A := L^{-1} A L^{-H}$ contain multiple chains of length 3 \cite{parikh2017}. Further examples include the fields of computer vision \cite{bronstein2016}, optimization \cite{straszak2015}, information theory \cite{albataineh2014, hejazi2016}, signal processing \cite{ding2016, nino2016, rao2015}, regularization \cite{noschese2016} and the simulation of power grids \cite{ronellenfitsch2015}.

With GMCP, the challenge shifts from finding the optimal parenthesization to
finding an optimal mapping to kernels. The standard algorithm operates on numbers that represent the chain. To solve GMCP, one has to work on symbolic expressions, rely on pattern matching and on the inference of properties.
In general, the solution to GMCP depends on
the available kernels. For the purpose of this paper, we  assume that the
kernels in $K$ offer the functionality necessary to guarantee that every input
chain is computable. In practice, the de-facto standard linear algebra
libraries BLAS and LAPACK offer such functionality, with kernels for matrix-matrix products and solve linear systems, with optionally transposed operands.

In typical chains, matrices often have properties and structure. This
information is relevant when trying to find the optimal way to compute an
expression; specialized, usually more efficient kernels can be used. This is
true for matrix products, but even more so for operations involving matrix
inversion. In Sec.~\ref{sec:properties} we explain that in order to fully take
advantage of properties and structure, not only must the GMC algorithm select
kernels based on this knowledge, it also needs to propagate and infer
knowledge along with intermediate results.

The MC algorithm finds the parenthesization that minimizes the FLOP count. In
practice, the number of FLOPs is not always an accurate metric to assess the
performance of linear algebra operations. More accurate results can be
achieved by taking the ``efficiency'' of the kernels into account.\footnote{It is
  often believed that the minimum execution time of an algorithm is attained
  by minimizing the number of floating point operations performed. This is not
  true, as not all flops cost the same (they are not equally ``efficient''.}
For this reason, our GMC algorithm allows to specify an arbitrary cost metric,
including vector measures, 
according to which the optimal solution is chosen.

In this paper, we describe GMCP, and present an al\-go\-rithm---the Generalized Matrix Chain (GMC) al\-go\-rithm---to solve it.

\paragraph{Organization of the paper}

The remainder of this section contains an overview of the terminology related
to GMCP, as well as a discussion of previous works. The standard dynamic programming
matrix chain algorithm is summarized in Sec~\ref{sec:mcalgorithm}.
Our GMC algorithm is presented in  Sec.~\ref{sec:gmcalgorithm}, 
and evaluated in Sec.~\ref{sec:results}. Finally, in Sec.~\ref{sec:conclusion} we discuss opportunities for future work.

\subsection{Terminology}
\label{sec:terminology}

\begin{description}
\item[Chain and Expression]
A valid input to the GMCP---a \emph{``matrix chain''} or simply
\emph{``chain''}---is a product $\mathcal{M} := f_0 \cdots f_{n-1}$ where
$f_i$ is a matrix or a vector that can be transposed and/or inverted. We use
$\mathcal{M}_{[ij]}$ to denote the product $f_i \cdots f_j$, which we call
\emph{``sub-chain''} or simply \emph{``expression''}. If $i=j$,
the chain consists of one single matrix and is denoted by
$\mathcal{M}_{[i]}$. Notice that the grammar in Fig.~\ref{fig:grammar} does not imply the correctness of expression, i.e. it does not guarantee that the dimensions of all operands match.

In this paper, we require the input to be a well-formed matrix chain of length two or higher. Vectors are considered to be matrices of size $n \times 1$ or $1 \times n$. Since scalars commute with matrices, we do not further consider them.
  	\item[Property] In addition to the specification shown in Fig.~\ref{fig:grammar}, we allow each symbol (i.e. each matrix) to be annotated with one or more properties. The grammar for the definition of the operands, including properties, is shown in Fig.~\ref{fig:grammarops}.
  \begin{figure}
    \begin{minipage}{\columnwidth} 
  	\begin{align*}
			\text{definitions} &\rightarrow \text{definition}^{+}\\
			\text{definition} &\rightarrow \textbf{Matrix} \; \textit{name} \; \text{size} \; \langle\text{property}^{*}\rangle\\
			\text{size} &\rightarrow (\textit{rows}, \textit{columns}) \\
			\text{property} &\rightarrow \textbf{LowerTriangular} \mid \textbf{Diagonal} \mid \ldots
	\end{align*}
    \end{minipage}
    \caption{Grammar describing the definition of operands.}
    \label{fig:grammarops}
  \end{figure}
  	Frequently encountered properties are: lower and upper triangular, symmetric, diagonal, and symmetric positive-definite (SPD).
      \item[Kernel]
        A kernel is an optimized routine for computing the solution of a
        well defined linear algebra problem, e.g.~{\tt C:=A*B}, {\tt C:=
            A$^\text{\tt -1}$*B}, as provided by libraries such as BLAS. Throughout this paper, we assume that there exists a set of
          specialized kernels that can take advantage of matrix
          properties.\footnote{As an example, consider the BLAS kernels GEMM,
            TRMM and SYMM, which all compute the product of two matrices. GEMM
            computes a general matrix-matrix product, while TRMM and SYMM
            require one operand to be triangular and symmetric,
            respectively. Compared to GEMM, TRMM and SYMM perform half of the scalar operations.}
	\item[Solution] A solution for the GMCP consists of a parenthesization
          of the input chain, in conjunction with a mapping of expressions to
          kernels. 
	\item[Cost function] This is a function to quantify the quality of a
          solution to the GMCP. Commonly used metrics are the total number of
          FLOPs, and the number of FLOPs per
          second, called \emph{performance}.\footnote{This use of
            \emph{performance} is, admittedly, a somewhat unfortunate choice
            of terminology adopted in the HPC community.} $\text{cost}(\mathcal{M})$ is used to denote the cost
          of computing the chain $\mathcal{M}$. 
	The solution that minimizes this cost function  is the optimal solution.
\end{description}

\subsection{Related Work}

The matrix chain problem is subject to a lot of research. The classic
algorithm to solve MCP uses dynamic programming and has $\mathcal{O}(n^3)$
complexity, where $n$ is the length of the chain \cite{cormen1990}. The best
known algorithm, by Hu and Shing, exploits the equivalence between MCP and the
triangulation of polygons to achieve $\mathcal{O}(n \; \text{log}(n))$
complexity \cite{hu1982, hu1984}. A number of approaches take parallelism into
account, some using multiple processors to reduce the time needed to find the
solution (which will be evaluated on a sequential system) \cite{ramanan1996,
  strate1990, bradford1998}, while others find an ordering that is optimal
when the matrix chain is evaluated on a parallel system
\cite{lee2003}. Nishida et al. present a version for GPUs
\cite{nishida2011}. Additionally, both sequential \cite{Chin1978} and parallel \cite{czumaj1996} algorithms exist that find approximate solutions. All the aforementioned algorithms deal with the basic problem of multiplying
matrices that are neither transposed nor inverted.

High-level languages such as Matlab allow to directly express instances of
GMCP, without explicit parenthesization, and link to  highly optimized
kernels. However, they put little to no effort into mapping the mathematical
problem to said kernels in a way that results in a highly efficient
evaluation. By constrast, expressions are typically evaluated according to
simple rules. For example, if the inverse operator is used in Matlab, then 
an inverse is computed explicitly, even though the mathematically equivalent solution
of solving a linear system is faster and numerically more stable; indeed, 
it is up to the user to rewrite the inverse in terms of the slash ({\tt/}) 
or backslash ({\tt\textbackslash}) operators, to ``enable'' the linear
systems. Furthermore, in Matlab products are always evaluated from left to right
\cite{matlabdoc}. Matrix properties are considered by inspecting matrix elements at runtime. Although not documented, Mathematica applies the same strategy to evaluate
expressions,\footnote{This can be easily tested by
  comparing the time necessary to evaluate $M_0 \cdots M_{k-1} x$ and
  $y M_0 \cdots M_{k-1}$, with $M_i \in \mathbb{R}^{n \times n}$,
  $x \in \mathbb{R}^{n \times 1}$ and $y \in \mathbb{R}^{1 \times n}$.}
including the choice between explicit inversion and solution of a linear
system. Recently, the Julia project~\cite{bezanson2012} set out to design a language that natively integrates
tools for scientific computing, including linear algebra; while high-level
expressions are accepted, they are evaluated just as in Matlab. Julia uses types to represent a small set of basic properties and uses multiple dispatch to select appropriate kernels.

An alternative approach consists in the use of (smart) expression templates in
C++, as employed by libraries such as Blaze \cite{iglberger2012a}, Blitz++
\cite{veldhuizen1998}, and Eigen \cite{eigenweb}. The main idea is to improve
performance by eliminating temporary operands and provide a domain-specific
language integrated within C++. However, similar to high-level languages,
expressions are evaluated according to very simple rules. To some extent, the
C++ library Armadillo is an exception \cite{sanderson2010}: It uses a simple algorithm which is not guaranteed to find the best solution to the MCP (the algorithm is discussed in detail in Sec.~\ref{sec:armadillo}).
Moreover, the choice regarding the treatment of the inverse operator is again left to the user. Similar to Julia, matrix properties are represented by types.

Recognizing that BLAS is not optimal across the full spectrum of operations
and problem sizes, some compilers such as Build to Order
(BTO) \cite{siek2008} and LGen \cite{spampinato2014, spampinato2016}
aim at generating directly code, without relying on standard build\-ing blocks.
BTO specializes in bandwidth-bound op\-er\-a\-tions (BLAS 1 and 2), while LGen
focuses on small-scale problems for which BLAS usually performs
poorly. Finally, one could approach the problem by systematically partitioning the input matrices, thus originating problems that fit exactly in a target cache level \cite{FabregatTraver:2011gu, fabregat-traver2011}.

In principle, GMCP (and thus also MCP) can be solved by means of a
search-based approach, as the one adopted in the linear algebra compiler
presented in \cite{fabregat-traver2013b, fabregat-traver2013a} (CLAK), which
exclusively relies on pattern matching. This approach has two drawbacks: The
type of pattern matching that CLAK uses is expensive, and due to the
search-based nature, the number of
explored solutions is exponential in the length of the chain,
even for the standard matrix chain problem \cite{cormen1990}.

\section{The Standard Matrix Chain Algorithm}
\label{sec:mcalgorithm}

MCP can be elegantly solved with a dynamic programming approach, both
in a top-down and a bottom-up fashion \cite{cormen1990}.
Here, we briefly explain the bottom-up
version, as it is the foundation for the algorithm presented in this paper.

Consider the chain $X:=ABCDE$ as an example. The algorithm proceeds by finding
the optimal parenthesization for parts of this chain of increasing length,
using the optimal solutions for sub-chains. Let us assume the algorithm
already computed all solutions for sub-chains of length up to three. The next
step consists of computing solutions for sub-chains of length
four. $\mathcal{M}_{[0,4]} = ABCDE$ has two such sub-chains,
$\mathcal{M}_{[0,3]} = ABCD$ and $\mathcal{M}_{[1,4]} = BCDE$. Let us
illustrate the step for $ABCD$:
There are three different ways to write this chain as a product of two shorter chains, or, to put it differently, three ways to split $\mathcal{M}_{[0,3]}$ into $\mathcal{M}_{[0,k]} \mathcal{M}_{[k+1,3]}$, namely for $k \in \{0, 1, 2\}$: $A(BCD)$, $(AB)(CD)$ and $(ABC)D$. The algorithm assigns a cost to all those products, and stores the best
solution together with its cost. The cost for $A(BCD)$ is the cost of computing
$\mathcal{M}_{[0,0]} = A$, plus the cost of $\mathcal{M}_{[1,3]} = BCD$, plus the cost of the product of $A$ and the result of $BCD$. The cost of $\mathcal{M}_{[0,0]}$ is known to be zero, and $\text{cost}(\mathcal{M}_{[1,3]})$ was already computed in a previous step because the length of $\mathcal{M}_{[1,3]}$ is three.
The same is done for $(AB)(CD)$, $(ABC)D$, as well as all possible ways to split $\mathcal{M}_{[1,4]}$. At this point, the algorithm uses all the results from the previous steps to find the best way to express $ABCDE$ as a product of two shorter parts.

The algorithm is shown in Fig.~\ref{fig:mcalgo}. The following arrays are used, where \code{solution} and \code{costs} have size $n \times n$, \code{sizes} is of size $n+1$:

\begin{description}
	\item[\code{solution}] The entry $\code{solution}[i][j]$ stores the
          integer $k$ which specifies the optimal split for $\mathcal{M}_{[i,j]}$. This array has the exact same role as the \emph{s} array in \cite{cormen1990}.
	
	\item[\code{costs}] The value of $\code{costs}[i][j]$ is the minimal
          cost for the computation of the sub-chain $\mathcal{M}_{[i,j]}$. The entries $\code{costs}[i][i]$ are initialized to $0$, while all other fields are initialized with $\infty$. This array has exactly the same role as the \emph{m} array in \cite{cormen1990}.
	\item[\code{sizes}] This array contains the operand sizes. $\code{sizes}[0]$ contains the number of rows of $\mathcal{M}_{[0]}$. For $i > 0$, $\code{sizes}[i]$ stores the number of columns of $\mathcal{M}_{[i-1]}$.
\end{description}

\begin{figure}
\hrule
\begin{lstlisting}{style=mystyle}
for $l \in \{1, \ldots, n-1\}$:
	for $i \in \{0, \ldots , n - l - 1\}$:
		$j := i + l$
		for $k \in \{i, \ldots, j-1\}$:
			c $:=$ 2*sizes[$i$]*sizes[$k+1$]*sizes[$j+1$]
			cost $:=$ costs[$i$][$k$] $+$ costs[$k+1$][$j$] $+$ c
			if cost < costs[$i$][$j$]:
				costs[$i$][$j$] $:=$ cost
				solution[$i$][$j$] $:=$ $k$
\end{lstlisting}
\hrule
\caption{The matrix chain algorithm.}
\label{fig:mcalgo}
\end{figure}

\section{The GMC Algorithm}
\label{sec:gmcalgorithm}

\subsection{Unary Operators}

\begin{figure}
\hrule
\lstset{numbers=left, xleftmargin=1.5em, framexleftmargin=1em, numbersep=0.5em}
\begin{lstlisting}
for $l \in \{1, \ldots, n-1\}$:
	for $i \in \{0, \ldots , n - l - 1\}$:
		$j := i + l$
		for $k \in \{i, \ldots, j-1\}$:
			expr $:=$ tmps[$i$][$k$] $\times$ tmps[$k+1$][$j$]
			kernel $:=$ match(expr) (*\label{line:match}*)
			cost $:=$ costs[$i$][$k$] $+$ costs[$k+1$][$j$] $+$ kernel.cost
			if cost < costs[$i$][$j$]:
				tmps[$i$][$j$] $:=$ create_tmp(expr) (*\label{line:createtmp}*)
				tmps[$i$][$j$].properties $:=$ infer_properties(expr) (*\label{line:inferprop}*)
				kernels[$i$][$j$] $:=$ kernel
				costs[$i$][$j$] $:=$ cost
				solution[$i$][$j$] $:=$ $k$
\end{lstlisting}
\hrule
\caption{The GMC algorithm.}
\label{fig:pseudocode}
\end{figure}

In its standard version, the matrix chain algorithm only works with
binary, non-commutative operators. To extend it to unary operators, we observe
that compositions of binary and unary operators on two operands can still be
seen as (an extended set of) binary operators. In fact, as long as it is
possible to assign a cost to those compositions of operations, the dynamic
programming approach remains applicable. The algorithm, however, becomes more
complex because in addition to the parenthesisation, it also has to identify 
which kernels can be applied and when.
To solve this problem, the GMC algorithm works on symbolic expressions, which
are represented as expression trees. Operands have a name, a size and a set of
properties (see Sec.~\ref{sec:properties}). Instead of the one-dimensional
array \code{sizes}, we now use the $n \times n$ array \code{tmps}, which is
used to store store symbolic temporary variables representing sub-chains.
In the following, consider the chain $\mathcal{M} = A^{-1} B C^{T}$ as an example. $\code{tmps}[i][j]$ contains the temporary that represents $\mathcal{M}_{[i,j]}$. The entry $\code{tmps}[i][i]$ is initialized with a symbolic representation of the matrix $\mathcal{M}_{[i]}$. For example $\code{tmps}[0][0]$ is $A^{-1}$. When the algorithm terminates, $\code{tmps}[1][2]$ contains a temporary $T_{12}$ that represents $B C^{T}$. The symbols representing those operands are used to create the expressions that have to be computed. For $i = j = 0$, $k = 2$, \code{expr} is the expression $\code{tmps}[0][0] \times \code{tmps}[1][2] = A^{-1} T_{12}$, which corresponds to the parenthesization $A^{-1} (B C^{T})$. New temporaries are created by the function \code{create\_tmp} (line \ref{line:createtmp}), which creates an operand with a unique name and correct sizes.

To select a suitable kernel, our algorithm relies on pattern matching as offered by MatchPy
\cite{krebber2017}, a Python library that implements discrimination nets, data structures for
efficient syntactic many-to-one pattern matching \cite{christian1993,
  graef1991, nedjah1997}. In many-to-one pattern matching, a set of patterns
and one expression are given, and it is tested whether or not any of those
patterns matches the expression. In our case, the set of patterns is the set
of kernels $K$. Some examples are shown in Table~\ref{tab:patterns}. In the
next section, we discuss the case in which more than one kernel matches the
target expression. %
\begin{table}
\centering
\caption{Examples of patterns for BLAS kernels.}
\begin{tabular}{llcc}
	Name & Pattern & Constraints & Cost\\\hline
	GEMM & $X Y$ & - & $2mnk$\\
	TRMM & $X Y$ & $\text{is\_lower\_triangular}(X)$ & $m^2 n$\\
	SYMM & $X Y$ & $\text{is\_symmetric}(X)$ & $m^2 n$\\
	TRSM & $X^{-1} Y$ &$\text{is\_lower\_triangular}(X)$ & $m^2 n$\\
	SYRK & $X^T X$ & - & $m^2 k$
\end{tabular}
\label{tab:patterns}
\end{table}
In the pseudocode in Fig.~\ref{fig:pseudocode}, pattern matching appears in line~\ref{line:match}.

To store the solution, in addition to the \code{solutions} array---which
contains the information on the parenthesization---it is necessary to also
keep track of the kernel used for the operation. In the MC algorithm, this is
not necessary because the kernel is always the same. For this purpose, we
introduce the $n \times n$ \code{kernels} array, whose entry $\code{kernels}[i][j]$ contains the kernel that is used to compute the temporary $\code{tmps}[i][j]$. In the end, the kernels are used to generate the code (see Sec.~\ref{sec:codegen}).
To simplify the discussion, here we only consider direct solvers for 
linear systems, using matrix factorizations if necessary. While the GMC
algorithm can also be applied to sparse linear algebra and iterative solvers,
the kernel selection becomes even more challenging \cite{nachtigal1992}.

\subsection{Properties}
\label{sec:properties}

Many linear algebra operations can be sped up by taking advantage of the
properties of the involved matrices. As a most basic example, the
multiplication of a lower triangular matrices with a full matrix requires $m^2
n$ scalar operations, as opposed to $2m^2 n$ operations for the multiplication
of two full matrices of the same sizes; likewise, a linear system $A^{-1} B$
where $A$ is symmetric positive definite can be solved faster than a system
where $A$ does not have any special properties.

Many properties are not mutually exclusive. As an example, a matrix can be banded and symmetric at the same time. Thus, it is possible that the same expression can be computed by
multiple kernels. Whenever more than one kernel matches (\code{match}
function, line \ref{line:match}), the algorithm selects the kernel that
minimizes the cost function (cost functions are discussed in the next section).

To fully take advantage of properties, it is certainly important to select the
best matching kernel, but it is even more critical to keep track of how
structure and properties propagate throughout the intermediate results, as different
kernels are applied. Take the product $A B^T$ as an example. If it is possible to assert that $B$ is upper triangular, $B^T$ is known to be lower triangular. Furthermore, the product of two lower triangular matrices is still lower triangular. Thus, if $A$ is lower triangular, the entire expression $A B^T$ has this property. Note that this property is independent of how $A B^T$ is computed, and it can be inferred without actually computing the result, solely by inspecting the symbolic expression.

This knowledge about properties can naturally be represented by inference rules, as for example
\begin{align*}
	\text{LoTri}(A) \land \text{LoTri}(B) &\rightarrow \text{LoTri}(AB), \\
	\text{LoTri}(A) \land \text{Diag}(B) &\rightarrow \text{LoTri}(AB), \\
	\text{LoTri}(A) &\rightarrow \text{UppTri}\left(A^T\right).
\end{align*}
In the GMC algorithm, the function \code{infer\_properties} (line \ref{line:inferprop}) is responsible for the inference of properties.
Intuitively, matrix properties are propagated from the bottom to the top of the expression tree. An example is shown in Fig.~\ref{fig:properties}.
	\newcommand{\drawmatrixsize}{0.8}
	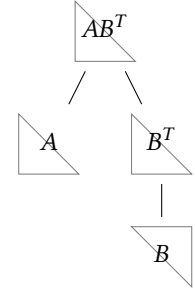
\begin{figure}
	\begin{tikzpicture}
		\node {\drawmatrix[lower, size=\drawmatrixsize]{$AB^T$}}
			child {node {\drawmatrix[lower, size=\drawmatrixsize]{$A$}}}
			child {node {\drawmatrix[lower, size=\drawmatrixsize]{$B^T$}}
				child {node {\drawmatrix[upper, size=\drawmatrixsize]{$B$}}}
			}
		;
	\end{tikzpicture}
	\caption{Example for the propagation of properties in $AB^T$ where $A$ is lower and $B$ is upper triangular.}
	\label{fig:properties}	
	\end{figure}
	This is done by recursively traversing the symbolic expression tree $\code{expr}$. In practice, this is implemented as a set of functions, with a dedicated function for each property. A part of the function \code{is\_lower\_triangular} is shown in Fig.~\ref{fig:islower}.
	
\begin{figure}
\hrule
\begin{lstlisting}
def is_lower_triangular(expr):
	if expr is Times:
		return $\forall \text{child} \in \text{expr.children}:$
		                     is_lower_triangular(child)
	if expr is Transpose:
		return is_upper_triangular(expr.child)
	if expr is Matrix:
		if $\text{LowerTriangular}$ in $\text{expr.properties}$:
			return True
		else:
			return False
	# ...

\end{lstlisting}
\hrule
\caption{Pseudocode implementation of the function is\_lower\_triangular.}
\label{fig:islower}
\end{figure}

Some languages, for example Matlab, do not infer properties symbolically but test for them by inspecting all entries of a matrix. The symbolic inference of properties has additional advantages over this approach: The cost is independent
of the matrix size. Furthermore, some properties might be masked by numerical
inaccuracies, which can have a considerable impact on subsequent computations:
A generalized eigenproblem $A x = \lambda B x$ is typically solved via a
reduction to a standard eigenproblem $A' y = \lambda y$; this is done by
computing the expression $A' := L^{-1} A L^{-T}$, where $L$ is lower
triangular and $A$ is symmetric; in floating point arithmetic, if $A'$ is computed by solving two linear
systems, symmetry is lost; thus, when computing the eigenvalues of $A'$, one
can only use a non-symmetric eigensolver, which is about three times more
expensive than a symmetric one~\cite{golub2013}, and worse yet, will deliver complex eigenvalues (with a very small imaginary part), doubling the amount of output data.

Let us consider the matrix chain $X := A^T A B$ as an example, where $A \in
\mathbb{R}^{n \times n}$ and $B \in \mathbb{R}^{n \times m}$ are dense
matrices with $n = 20$ and $m = 15$, using the FLOP count as a metric. The
first possible solution is to compute $X$ via two general matrix-matrix products:
\begin{align*}
	W&:= AB \text{,}\\
	X&:= A^T W \text{.}
\end{align*}
In this case, $W$ is of size $n \times m$ and has no special properties. Both products require $2n^2k = 12000$ FLOPs; the overall cost of this solution is $24000$ FLOPs. The alternative solution results in an intermediate $W \in \mathbb{R}^{n \times n}$ that is symmetric and positive definite:
\begin{align*}
	W&:= A^T A \text{,}\\
	X&:= W B \text{.}
\end{align*}
If $A^T A$ is computed as a general matrix-matrix product, the cost is $2n^3 =
16000$ FLOPs. Disregarding that $W$ is symmetric and computing $WB$ as another
general matrix-matrix product, we get $2n^2 k = 12000$ additional FLOPs. By
taking advantage of the property instead, one computes this product with half
the number of FLOPs ($n^2 k = 6000$). Hence, the solution obtained by using a
specialized kernel has a cost of $22000$ FLOPs, compared to $28000$
FLOPs.
This example shows that properties not only lead to better solutions, but also
to solutions that might differ in the parenthesization. Note: instead of
computing $A^T A$ as a general matrix-matrix product, it is also 
possible to use the specialized SYRK kernel, performing half the number of FLOPs.

\subsection{Cost Functions}
\label{sec:costfunction}

It is well understood that properties such as temporal and spacial locality
impact the execution time of an algorithm as much as, or even more than the
number of FLOPs~\cite{dongarra1990}. In fact, two solutions that are identical in terms of FLOPs might have very different actual execution times; it might even be the case that the faster algorithm performs significantly more FLOPs than the slower one. One such example is the reduction of a symmetric matrix to tridiagonal form \cite{golub2013}. As a second example, consider the matrix chain $ABCDE$ with matrix sizes (from left to right) $130$, $700$, $383$, $1340$, $193$ and $900$. The parenthesization that results in the smallest number of FLOPs is $(((AB)C)D)E$ with $3.16 \times 10^8$ FLOPs. The parenthesization that results in the shortest execution time is $((AB)(CD))E$, even though the number of FLOPs is slightly higher with $3.32 \times 10^8$. With an execution time of $7.6$ milliseconds, the first parenthesization is about 10\% slower than the second parenthesization, which takes $6.8$ milliseconds.\footnote{The presented times are the minimum out of 100 repetitions, using BLAS wrappers in Julia $0.6.1$ on an Intel Core i5 with 2,7 GHz.}
That said, the FLOP count is still the most commonly used metric in practice; this is certainly because of its simplicity, but also because the performance modeling and prediction of linear algebra kernels remains a challenging, largely unsolved problem~\cite{peise2012}. 

We want the GMC algorithm to use an arbitrary metric. To take into account the different operations, a cost function has to be provided that is defined for the different kernels. Once a kernel is identified, the function is used to determine its cost. This is done based on the sizes of the operands and their properties.

A more accurate metric than the FLOP count is the performance (in
FLOPs/sec); if the execution time of
the matrix chain algorithm is of no concern (Note: this is the time it takes
to find the optimal mapping, and not the execution time of the resulting
algorithm), real measurements could be used, for example using performance
modeling tools such as ELAPS \cite{peise2015}. However, these approaches are
still not able to accurately predict the performance of an entire chain. In
fact, performance is not composable, which means that the
combination of the performance of two kernels executed separately will not be
the same as the performance of the same kernels executed back to
back~\cite{peise2015a}. One of the reasons for this is the state of the
cache. Despite this hurdle, performance is still a better approximation than the number of FLOPs.

A cost function can also take into account accuracy: The explicit inversion of matrices should be avoided if it is possible to solve a linear system instead, both for performance and stability reasons. Since explicit inversion is more expensive, using a cost metric based on performance automatically leads to a solution that favors the solution of linear systems.

\subsection{Complexity and Completeness}

The functions used in the GMC algorithm and their complexity are described below. For considerations regarding the time complexity, it is important to note that the size of the expression tree representing \code{expr} is limited. The most complex expressions have the form $f_1(A) \cdot f_2(B)$, with $f$ being the transposition, inversion, or the combination of both, and operands $A$ and $B$. Thus, those trees have at most five nodes and three levels. The same is true for the size of the patterns, as kernels that compute more complex expressions than $f_1(A) \cdot f_2(B)$ are not applicable.

\begin{description}
	\item[match] The complexity of syntactic pattern matching with discrimination nets does not depend on the number of patterns and is bounded by the size of the patterns, which in our case is constant. It follows that the complexity of pattern matching is $\mathcal{O}(1)$.
	\item[create\_tmp] This function creates a symbolic temporary matrix that represents the result of computing $\mathcal{M}_{[i,j]}$. For example, for an outer product $a b^T$, with $a \in \mathbb{R}^n$ and $b \in \mathbb{R}^m$, a temporary matrix $T \in \mathbb{R}^{n \times m}$ will be created. This function creates a symbolic object with a unique name and correct sizes. The size is determined by traversing the expression tree, that is bounded in its size by a constant, so this function is in $\mathcal{O}(1)$.
	\item[infer\_properties] Since the size of the expression trees is limited by a small constant, this function has a complexity of $\mathcal{O}(p)$, where $p$ is the number of properties. Those functions are then also used for the constraints of the patterns that represent kernels (Table~\ref{tab:patterns}).
\end{description}

The loop body is executed $\mathcal{O}(n^3)$ times, where $n$ is the length of the matrix chain (see \cite{cormen1990}). Thus, the complexity of the entire algorithm is $\mathcal{O}(n^3 + n^3 p)\text{.}$ This could be further reduced to $\mathcal{O}(n^3 + n^2p)$ by inferring properties outside of the $k$ loop only for the temporary that might be used in the solution.

\paragraph{Completeness}

We stress that the GMC algorithm might deliver a solution 
even if one or more sub-chains are not computable because
no suitable kernel is found.
Let us assume we are given the matrix chain $X := A^{-1} B^{-1} C$, and there is no kernel that computes $X^{-1} Y^{-1}$, so $A^{-1} B^{-1}$ can not be computed. In this case, find\_sequence would return $\infty$ as the cost of $A^{-1} B^{-1}$. However, this chain can still be computed by solving two linear systems:
\begin{align*}
	T &:= B^{-1} C \\
	X &:= A^{-1} T \text{.}
\end{align*}
In general, the GMC algorithm will find a solution if there is at least one parenthesization such that all exposed binary operations can be computed.

\subsection{Code Generation}
\label{sec:codegen}

\begin{figure}
\hrule
\begin{lstlisting}
def construct_solution($i$, $j$):
	if $i \neq j$:
		yield from construct_solution($i$, solution[$i$][$j$])
		yield from construct_solution(solution[$i$][$j$]$+1$, $j$)
		yield kernels[$i$][$j$]
\end{lstlisting}
\hrule
\caption{Function to construct the solution. \code{yield} and \code{yield from} behave as the corresponding Python keywords.}
\label{fig:constructsolution}
\end{figure}

Retrieving the sequence of kernels that was identified as the optimal solution
is done by calling \code{con\-struct\_so\-lu\-tion}$(0, n-1)$, where $n$ is the
length of the chain. The function \code{con\-struct\_so\-lu\-tion} is show in
Fig.~\ref{fig:grammarops}. The kernels are returned in an order that respects
dependencies. However, in some cases, kernel calls can be reordered. This is
for example the case for the chain $(AB)(CD)$, where $AB$ and $CD$ can be
computed independently. Since performance is not composable, different orders likely result in different performance; the
prediction of the best ordering is again a difficult task and it could be
added as a final optimization.

\section{Results}
\label{sec:results}

To evaluate the quality of the algorithms generated by the GMC algorithm, we compare against
Julia\footnote{Development version of Julia 0.7 from September 4, 2017.},
Matlab\footnote{Version 2017a.},
Eigen\footnote{Version 3.3.4.},
Blaze\footnote{Development version of Blaze 3.3 from September 4, 2017.}
and Armadillo\footnote{Version 7.960.2.}. We link against the Intel MKL implementation of BLAS and LAPACK (MKL 2017 update 3) \cite{mkldoc}, with the exception of
Matlab, which instead uses LAPACK 3.5.0 and MKL 11.3.1.
The GMC algorithm generates Julia code that uses BLAS and LAPACK
wrappers.
As a cost metric, FLOPs are used. When possible, we consider two different implementations for each library and language: \texttt{naive} and \texttt{recommended}. The \texttt{naive} implementation is the one that comes closest to the
mathematical description of the problem. As an example, in Julia $A^{-1} B$ is
implemented as \texttt{inv(A)*B}. However, since the documentations almost
always discourage this use of the inverse operator, we also consider a so
called \texttt{recommended} implementation, which uses dedicated functions to
solve linear systems (\texttt{A\textbackslash B}).

In the following, we describe the different implementations. As examples, in Table~\ref{tab:implementations} we provide the implementations of $A^{-1} B C^T$ where $A$ is symmetric positive definite and $C$ is lower triangular.

\begin{table}
\centering
\caption{Implementations of $A^{-1} B C^T$.}
\begin{tabular}{lp{6.1cm}}
	Name & Implementation \\\hline
	GMC & \texttt{trmm!('R', 'L', 'T', 'N', 1.0, C, B)
    posv!('L', A, B)
} \\
	Jl n & \texttt{inv(A)*B*C'} \\
	Jl r & \texttt{(A\textbackslash B)*C'} \\
	Arma n & \texttt{arma::inv\_sympd(A)*B*(C).t()} \\
	Arma r & \texttt{arma::solve(A, B)*C.t()} \\
	Eig n & \texttt{A.inverse()*B*C.transpose()} \\
	Eig r & \texttt{A.llt().solve(B)*C.transpose()} \\
	Bl n & \texttt{blaze::inv(A)*B*blaze::trans(C)} \\
	Mat n & \texttt{inv(A)*B*C'} \\
	Mat r & \texttt{(A\textbackslash B)*C'} \\
\end{tabular}
\label{tab:implementations}
\end{table}

\begin{description}
	\item[Julia] Properties are expressed via types. The \texttt{naive}
          implementation uses \texttt{inv()}, while the recommended one uses
          the \texttt{\slash} and \textbf{\textbackslash} operators.
	\item[Matlab] The \texttt{naive} implementation uses \texttt{inv()}, the \texttt{rec\-om\-mend\-ed} the \texttt{\slash} and \textbf{\textbackslash} operators.
	\item[Eigen] The \texttt{recommended} implementation uses the recommended linear systems solvers based on the matrix properties, as well as views to describe properties.
	\item[Armadillo] In the \texttt{naive} implementation, specialized
          functions are used for the inversion of SPD and diagonal
          matrices. For \texttt{solve}, we use the \texttt{solve\_opts::fast}
          option to disable an expensive refinement. In addition,
          \texttt{trimatu} and \texttt{trimatl} are used for triangular matrices.
	\item[Blaze] Since Blaze does not offer functions to solve linear systems, there is no \texttt{recommended} implementation. Properties are specified by adaptors.
\end{description}

The example problems are generated randomly, to include a mix of square and rectangular matrices as well as vectors. The length of the chains is chosen uniformly in the range $[3, 10]$. Matrices can be transposed and/or inverted, and may have one of the following properties: Diagonal, lower triangular, upper triangular,
symmetric, symmetric positive definite.
Matrix sizes are chosen uniformly from between $50$ and $2000$ in steps of $50$. For the experiments, we use $100$ chains.
The measurements were taken on an Intel Xeon E5-2680 v3 with 2.5 GHz and 64 GB RAM.
All experiments were repeated 20 times; and the average is reported.

The GMC algorithm takes on average 0.03s to generate solutions, and in all cases less than 0.07s. Thus, it would even be possible to use the GMC algorithm in an interactive environment. Notice that the generation time does not depend on matrix sizes.

\pgfplotstableread[col sep=comma]{speedup_mean.csv}\speedup

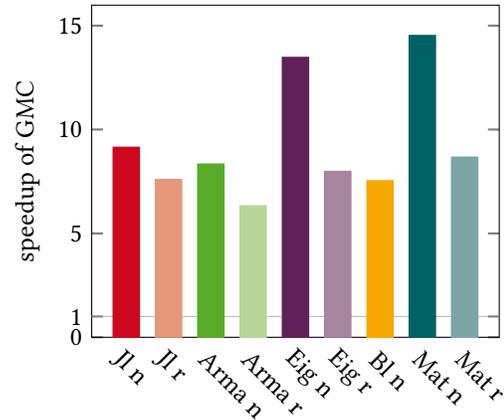
\begin{figure}
\centering
\begin{tikzpicture}
\begin{axis}[
	width=7cm,
	height=6cm,
	ybar = 6pt,
	ymin=0,
	xtick=\empty,
	enlarge x limits=0.15,
	xticklabels={Jl n, Jl r, Arma n, Arma r, Eig n, Eig r, Bl n, Mat n, Mat r},
	xtick={-0.84,-0.63,...,1.2},
	x tick style={
		draw=none,
	},
    x tick label style={
		xshift=-5pt,
		rotate=-45,
		anchor=west,
		},
	ylabel={speedup of GMC},
	tick style={line width=0.8pt},
	extra y ticks={1},
	extra tick style={
		grid=major,
	},
	legend style={
		at={(0.5,-0.05)},
    	anchor=north,
    	legend columns=5,
    	draw=none,
    	legend cell align=left,
    },
	]
	\addplot[draw=rwthred, fill=rwthred] table[y=naive_julia] {\speedup};
	\addplot[draw=rwthred50, fill=rwthred50] table[y=recommended_julia] {\speedup};
    \addplot[draw=rwthgreen, fill=rwthgreen] table[y=naive_armadillo] {\speedup};
    \addplot[draw=rwthgreen50, fill=rwthgreen50] table[y=recommended_armadillo] {\speedup};
    \addplot[draw=rwthviolet, fill=rwthviolet] table[y=naive_eigen] {\speedup};
    \addplot[draw=rwthviolet50, fill=rwthviolet50] table[y=recommended_eigen] {\speedup};
    \addplot[draw=rwthorange, fill=rwthorange] table[y=naive_blaze] {\speedup};
    \addplot[draw=rwthpetrol, fill=rwthpetrol] table[y=naive_matlab] {\speedup};
    \addplot[draw=rwthpetrol50, fill=rwthpetrol50] table[y=recommended_matlab] {\speedup};

\end{axis}
\end{tikzpicture}\vspace{-0.3cm}
\caption{Average speedup of the GMC-generated code over other libraries and languages.}
\label{fig:speedup}
\end{figure}

The average speedup of the GMC-generated code over the other libraries and languages is between 6 and 15, as shown in Fig.~\ref{fig:speedup}. One can observe that the execution times of Julia, Armadillo and Blaze are comparable. The naive implementations in Eigen and Matlab are noticeably slower. As expected, the recommended implementations perform better than the corresponding naive implementations. In general, Armadillo emerges as the second fastest solution. This is likely because Armadillo, unlike all the other systems, uses a heuristic to finde better solutions for the matrix chain problem.

\paragraph{Matrix Chains in Armadillo}
\label{sec:armadillo}

As mention before, Ar\-madil\-lo is the only system that considers the matrix chain problem to some extent, using a simplified algorithm to solve it. For a chain $ABCD$, $(ABC)D$ is chosen if $ABC$ is smaller in size than $BCD$. Otherwise, $A(BCD)$ is used. Similarly, for a chain $ABC$, either $(AB)C$ or $A(BC)$ is chosen, depending on the sizes of $AB$ and $BC$. Chains with more than four matrices are broken down into chains of length $n \leq 4$. This happens in a deterministic way that depends on how expression templates are constructed in Armadillo. Using this method, not all parenthesizations can be found; $(AB)(CD)$ is not possible.
However, parenthesizations found by this algorithm have the advantage that they have good caching behavior: Every binary product uses the result of the previous one. As an example, consider $A((BC)D)$, which results in the following sequence of kernels:%
\begin{align*}
	T_1 &:= BC \\
	T_2 &:= T_1 D \\
	T_3 &:= A T_2
\end{align*}

\pgfplotstableread[col sep=comma]{execution_time.csv}\executiontime

\pgfplotstablesort[sort key=generated0,sort cmp=float <]\executiontimesorted{\executiontime}

\newcommand{\markscaling}{1}
\newcommand{\markscalingsmall}{0.7}

\begin{figure*}[]
    \centering
    \begin{tikzpicture}
        \begin{axis}[
            height=7cm,
            width=\textwidth,
            xlabel={Test problems},
            ylabel={Execution time [s]},
            grid=major,
            xtick=\empty,
            ymode=log,
            enlarge x limits=0.03,
            enlarge y limits=0.05,
            legend style={
            	at={(0.79,0.22)},
            	anchor=north,
            	column sep=0.2em,
            },
            legend columns=5,
        ]
            \addplot[
            	fill=rwthblue,
            	draw=rwthblue,
            	mark=+,
            	only marks,
            	mark options={scale=\markscaling},
            ]
                table[
                    x expr=\coordindex,
                    y=generated0,
                ] {\executiontimesorted};
            \addlegendentry{GMC};
            \addplot[
            	fill=rwthred,
            	draw=rwthred,
            	mark=triangle*,
            	only marks,
            	mark options={scale=\markscaling}
            ]
                table[
                    x expr=\coordindex,
                    y=naive_julia,
                ] {\executiontimesorted};
            \addlegendentry{Jl n};
            \addplot[
            	fill=rwthred50,
            	draw=rwthred50,
            	mark=triangle,
            	only marks,
            	mark options={scale=\markscaling}
            ]
                table[
                    x expr=\coordindex,
                    y=recommended_julia,
                ] {\executiontimesorted};
            \addlegendentry{Jl r};
            \addplot[
            	fill=rwthgreen,
            	draw=rwthgreen,
            	mark=square*,
            	only marks,
            	mark options={scale=\markscalingsmall}
            ]
                table[
                    x expr=\coordindex,
                    y=naive_armadillo,
                ] {\executiontimesorted};
            \addlegendentry{Arma n};
            \addplot[
            	fill=rwthgreen50,
            	draw=rwthgreen50,
            	mark=square,
            	only marks,
            	mark options={scale=\markscalingsmall}
            ]
                table[
                    x expr=\coordindex,
                    y=recommended_armadillo,
                ] {\executiontimesorted};
            \addlegendentry{Arma r};
            \addplot[
            	fill=rwthviolet,
            	draw=rwthviolet,
            	mark=diamond*,
            	only marks,
            	mark options={scale=\markscaling}
            ]
                table[
                    x expr=\coordindex,
                    y=naive_eigen,
                ] {\executiontimesorted};
            \addlegendentry{Eig n};
            \addplot[
            	fill=rwthviolet50,
            	draw=rwthviolet50,
            	mark=diamond,
            	only marks,
            	mark options={scale=\markscaling}
            ]
                table[
                    x expr=\coordindex,
                    y=recommended_eigen,
                ] {\executiontimesorted};
            \addlegendentry{Eig r};
            \addplot[
            	fill=rwthorange,
            	draw=rwthorange,
            	mark=asterisk,
            	only marks,
            	mark options={scale=\markscaling}
            ]
                table[
                    x expr=\coordindex,
                    y=naive_blaze,
                ] {\executiontimesorted};
            \addlegendentry{Bl n};
            \addplot[
            	fill=rwthpetrol,
            	draw=rwthpetrol,
            	mark=*,
            	only marks,
            	mark options={scale=\markscalingsmall}
            ]
                table[
                    x expr=\coordindex,
                    y=naive_matlab,
                ] {\executiontimesorted};
            \addlegendentry{Mat n};
            \addplot[
            	fill=rwthpetrol50,
            	draw=rwthpetrol50,
            	mark=o,
            	only marks,
            	mark options={scale=\markscalingsmall}
            ]
                table[
                    x expr=\coordindex,
                    y=recommended_matlab,
                ] {\executiontimesorted};
            \addlegendentry{Mat r};
        \end{axis}
    \end{tikzpicture}
    \caption{Execution times of all test problems (sorted by the execution time of the GMC-generated code).}\label{fig:executiontime}
\end{figure*}
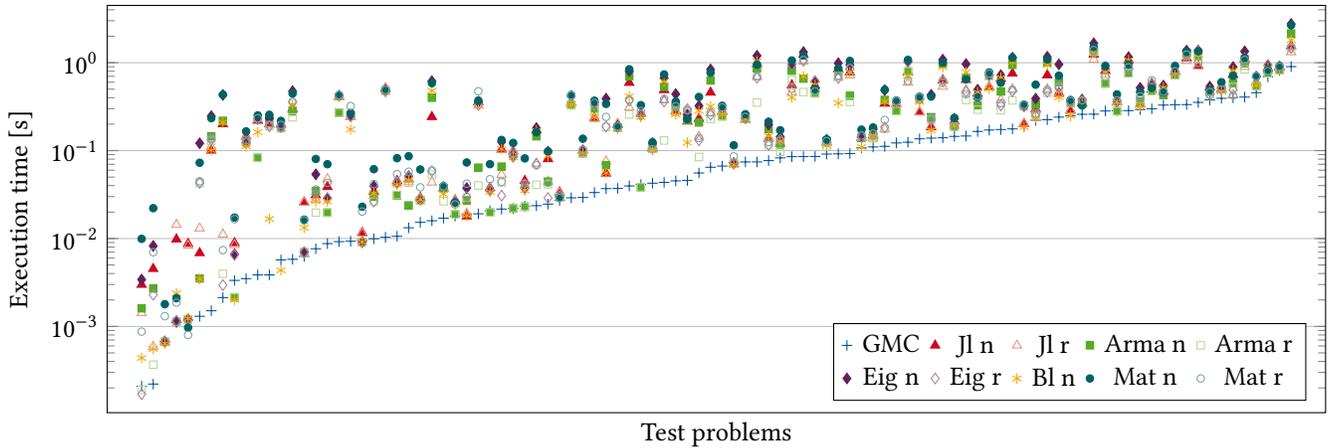

The execution times of all experiments are shown in Fig.~\ref{fig:executiontime}. The average multiplication time of the GMC solutions is 0.13s, ranging from 0.0002s to 0.9s. For 86\% of the test cases, the code generated by the GMC algorithm is the fastest. In those cases when the GMC implementation is not the fastest, it is never more than a factor of 1.66 worse than the best solution. In only 4\% of the test cases, other solutions are more than 1.1 times faster than the generated code.
In at least 10\% of the test cases (from a minimum of 10\% for Armadillo \texttt{recommended} to 25\% for Eigen \texttt{naive}), the other implementations are more than 10 times slower than the GMC solutions. In the worst case, the naive Eigen and Matlab solutions are about 200 times slower than the best solution. The maximal speedup of GMC implementations over other solutions depends both on the length of the chains and matrix sizes. Since it is possible for the GMC algorithm to find solutions with lower asymptotic complexity, the speedup of GMC implementations can potentially become arbitrarily large.

Inspecting the cases where the GMC-generated code is not the fastest reveals some patterns. There are multiple cases with chains of the form $M_1 \cdots M_{n-1} M_n v_1 v_2^T$, where $v_1$ and $v_2$ are vectors. The best algorithm (in terms of FLOPs) for this type of chain is usually to first compute a sequence of matrix-vector products
%
\begin{align*}
	t_1 &:= M_n v_1 \\
	t_2 &:= M_{n-1} t_1 \\
	 & \ldots \\
	t_n &:= M_{1} t_{n-1}
\end{align*}
and then compute the outer product $t_n v_2^T$. This is the solution that the GMC algorithms finds, but also the one used by Armadillo, Blaze and Eigen. In Armadillo, this sequence is found by the heuristic for matrix chains. While Blaze does not solve the matrix chain problem, products of the form $A B v$, where $v$ is a vector, are computed as $A (B v)$ \cite{iglberger2012a}, resulting in the same sequence of kernels. For a chain $M_1 M_2^{-1} v_1 v_2^T$, the recommended Eigen implementation performs well because of the inverse operator: Using the \texttt{.solve()} method in Eigen to solve the linear system result in the parenthesization $(M_1 (M_2^{-1} v_1)) v_2^T$. Measuring the execution time of the individual kernels reveals that Armadillo, Blaze and Eigen have an implementation of an outer product that is significantly faster than the BLAS implementation used in the GMC implementation.

For all remaining test cases where other solutions are more than 5\% faster than the GMC-generated code, the evaluation from left to right happens to be optimal (or almost optimal) in terms of FLOPs. As a result, all implementations use the same (or comparable) parenthesizations. Thus, in can be concluded that other implementations outperform the GMC-generated code because the respective languages and libraries use faster implementations of some kernels, or because the GMC-generated code contains some overhead. This also means that for the presented test cases, the cost function is sufficiently accurate to find good solutions.

\section{Conclusion and Future Work}
\label{sec:conclusion}

In this paper, we introduce a number of extensions to the standard matrix chain algorithm to generate efficient algorithms and code for classes of problems as they commonly occur in actual applications.
The extensions include additional operations like transposition and solution of linear systems, the use of matrix properties, which are necessary to take
advantage of specialized kernels and a flexible cost function. Our experimental results provide evidence that the algorithms generated by the GMC algorithm substantially outperform existing libraries and languages.
  
In summary, this paper makes the following contributions.
\begin{itemize}
\item[-] In the MCP, expressions only consist of products. We
  demonstrate how the original algorithm can be extended to deal
  with more complex expressions, which also involve unary operators, by treating
  compositions of unary and binary operators as an extended set of binary
  operators. 
\item[-] We discuss the use of properties and the importance of propagating
  them to automatically map linear algebra expressions to sequences of
  specialized kernels that can take advantage of said properties.
\item[-] We provide evidence that the GMC algorithm is a useful tool for the generation of efficient code for practical linear algebra problems. In addition to being part of a compiler for linear algebra, it could even be used in interactive environments such as Julia.
\end{itemize}

For the linear algebra compiler Linnea \cite{barthels2016src}, the GMC algorithm allows to find good solutions while at the same time keeping the search space relatively small.

The GMC algorithm can be extended even further in multiple ways. For the
purpose of this paper, we assumed that a kernel for $X:= A^{-1} B^{-1}$ is
provided. In practice, such kernels do not
exists. Instead of manually constructing them from existing BLAS and LAPACK kernels, it would also be possible to again
use Linnea to generate them when necessary, as this is the exact type of problem that Linnea solves. Of course, in that case one can easily expand the set of operations even further, adding unary operators like complex conjugation, matrix logarithm or exponentiation and other non-commutative binary operators.

In general, the metric does not have to be a measure of the execution time; it can be a measure of numerical accuracy, memory consumption, number of bytes moved, or a combination of multiple objectives. It is also possible to use a vector for the metric, as long as addition and a total ordering is defined on the vector space.

\begin{acks}
Financial support from the Deutsche Forschungsgemein\-schaft (German Research Foundation) through grants GSC 111 and BI 1533/2-1 is gratefully acknowledged. We thank Ryan Curtin for helpful discussions on how the matrix chain problem is solved in Armadillo. We thank Sadulla Aghayev and Edilbert Christhuraj for providing the example in Sec.~\ref{sec:costfunction}.
\end{acks}


\bibliographystyle{ACM-Reference-Format}
\bibliography{../../shared_tex_files/PhD_bibliography,../../shared_tex_files/PhD_bibliography_papers,../../shared_tex_files/my_publications}

\end{document}

%% file: RWTH_colors_v2.tex
\definecolor{rwthblue}   {RGB}{  0,  84, 159}
\definecolor{rwthblue75}{RGB}{ 64, 127, 183}
\definecolor{rwthblue50}{RGB}{142, 186, 229}
\definecolor{rwthblue25}{RGB}{199, 221, 242}
\definecolor{rwthblue10}{RGB}{232, 241, 250}

\definecolor{rwthblack}  {RGB}{  0,   0,   0}
\definecolor{rwthblack75}{RGB}{100, 101, 103}
\definecolor{rwthblack50}{RGB}{156, 158, 159}
\definecolor{rwthblack25}{RGB}{207, 209, 210}
\definecolor{rwthblack10}{RGB}{236, 237, 237}

\definecolor{rwthdarkgray}{named}{rwthblack75}
\definecolor{rwthgray}{named}{rwthblack50}
\definecolor{rwthlightgray}{named}{rwthblack25}
\definecolor{rwthverylightgray}{named}{rwthblack10}

\definecolor{rwthmagenta}  {RGB}{227,   0, 102}
\definecolor{rwthmagenta75}{RGB}{233,  96, 136}
\definecolor{rwthmagenta50}{RGB}{241, 158, 177}
\definecolor{rwthmagenta25}{RGB}{249, 210, 218}
\definecolor{rwthmagenta10}{RGB}{253, 238, 240}

\definecolor{rwthyellow}  {RGB}{255, 237,   0}
\definecolor{rwthyellow75}{RGB}{255, 240,  85}
\definecolor{rwthyellow50}{RGB}{255, 245, 155}
\definecolor{rwthyellow25}{RGB}{255, 250, 209}
\definecolor{rwthyellow10}{RGB}{255, 253, 238}

\definecolor{rwthpetrol}   {RGB}{  0,  97, 101}
\definecolor{rwthpetrol75}{RGB}{ 45, 127, 131}
\definecolor{rwthpetrol50}{RGB}{125, 164, 167}
\definecolor{rwthpetrol25}{RGB}{191, 208, 209}
\definecolor{rwthpetrol10}{RGB}{230, 236, 236}

\definecolor{rwthturquoie}   {RGB}{  0, 152, 161}
\definecolor{rwthturquoise75}{RGB}{  0, 177, 183}
\definecolor{rwthturquoise50}{RGB}{137, 204, 207}
\definecolor{rwthturquoise25}{RGB}{202, 231, 231}
\definecolor{rwthturquoise10}{RGB}{235, 246, 246}

\definecolor{rwthgreen}  {RGB}{ 87, 171,  39}
\definecolor{rwthgreen75}{RGB}{141, 192,  96}
\definecolor{rwthgreen50}{RGB}{184, 214, 152}
\definecolor{rwthgreen25}{RGB}{221, 235, 206}
\definecolor{rwthgreen10}{RGB}{242, 247, 236}

\definecolor{rwthlightgreen}   {RGB}{189, 205,   0}
\definecolor{rwthlightgreen75}{RGB}{208, 217,  92}
\definecolor{rwthlightgreen50}{RGB}{224, 230, 154}
\definecolor{rwthlightgreen25}{RGB}{240, 243, 208}
\definecolor{rwthlightgreen10}{RGB}{249, 250, 237}

\definecolor{rwthorange}  {RGB}{246, 168,   0}
\definecolor{rwthorange75}{RGB}{250, 190,  80}
\definecolor{rwthorange50}{RGB}{253, 212, 143}
\definecolor{rwthorange25}{RGB}{254, 234, 201}
\definecolor{rwthorange10}{RGB}{255, 247, 234}

\definecolor{rwthred}  {RGB}{204,   7,  30}
\definecolor{rwthred75}{RGB}{216,  92,  65}
\definecolor{rwthred50}{RGB}{230, 150, 121}
\definecolor{rwthred25}{RGB}{243, 205, 187}
\definecolor{rwthred10}{RGB}{250, 235, 227}

\definecolor{rwthbordeaured}   {RGB}{161,  16,  53}
\definecolor{rwthbordeauxred75}{RGB}{182,  82,  86}
\definecolor{rwthbordeauxred50}{RGB}{205, 139, 135}
\definecolor{rwthbordeauxred25}{RGB}{229, 197, 192}
\definecolor{rwthbordeauxred10}{RGB}{245, 232, 229}

\definecolor{rwthviolet}  {RGB}{ 97,  33,  88}
\definecolor{rwthviolet75}{RGB}{131,  78, 117}
\definecolor{rwthviolet50}{RGB}{168, 133, 158}
\definecolor{rwthviolet25}{RGB}{210, 192, 205}
\definecolor{rwthviolet10}{RGB}{237, 229, 234}

\definecolor{rwthpurple}  {RGB}{122, 111, 172}
\definecolor{rwthpurple75}{RGB}{155, 145, 193}
\definecolor{rwthpurple50}{RGB}{188, 181, 215}
\definecolor{rwthpurple25}{RGB}{222, 218, 235}
\definecolor{rwthpurple10}{RGB}{242, 240, 247}

%% file: CGO_2018_arXiv.bbl

\begin{thebibliography}{00}


\ifx \showCODEN    \undefined \def \showCODEN     #1{\unskip}     \fi
\ifx \showDOI      \undefined \def \showDOI       #1{#1}\fi
\ifx \showISBNx    \undefined \def \showISBNx     #1{\unskip}     \fi
\ifx \showISBNxiii \undefined \def \showISBNxiii  #1{\unskip}     \fi
\ifx \showISSN     \undefined \def \showISSN      #1{\unskip}     \fi
\ifx \showLCCN     \undefined \def \showLCCN      #1{\unskip}     \fi
\ifx \shownote     \undefined \def \shownote      #1{#1}          \fi
\ifx \showarticletitle \undefined \def \showarticletitle #1{#1}   \fi
\ifx \showURL      \undefined \def \showURL       {\relax}        \fi
\providecommand\bibfield[2]{#2}
\providecommand\bibinfo[2]{#2}
\providecommand\natexlab[1]{#1}
\providecommand\showeprint[2][]{arXiv:#2}

\bibitem[\protect\citeauthoryear{??}{mkl}{2017}]%
        {mkldoc}
 \bibinfo{year}{2017}\natexlab{}.
\newblock \bibinfo{title}{{I}ntel\textregistered {M}ath {K}ernel {L}ibrary
  documentation}.
\newblock
  \bibinfo{howpublished}{\url{https://software.intel.com/en-us/mkl-reference-manual-for-c}}.
    (\bibinfo{year}{2017}).
\newblock


\bibitem[\protect\citeauthoryear{??}{mat}{2017}]%
        {matlabdoc}
 \bibinfo{year}{2017}\natexlab{}.
\newblock \bibinfo{title}{{M}atlab documentation}.
\newblock \bibinfo{howpublished}{\url{http://www.mathworks.com/help/matlab}}.
  (\bibinfo{year}{2017}).
\newblock


\bibitem[\protect\citeauthoryear{Albataineh and Salem}{Albataineh and
  Salem}{2014}]%
        {albataineh2014}
\bibfield{author}{\bibinfo{person}{Zaid Albataineh} {and}
  \bibinfo{person}{Fathi~M. Salem}.} \bibinfo{year}{2014}\natexlab{}.
\newblock \showarticletitle{{A} {B}lind {A}daptive {CDMA} {R}eceiver {B}ased on
  {S}tate {S}pace {S}tructures}.
\newblock \bibinfo{journal}{{\em CoRR\/}}  \bibinfo{volume}{abs/1408.0196}
  (\bibinfo{year}{2014}).
\newblock


\bibitem[\protect\citeauthoryear{Anderson, Bai, Bischof, Blackford, Dongarra,
  Du~Croz, Greenbaum, Hammarling, McKenney, and Sorensen}{Anderson
  et~al\mbox{.}}{1999}]%
        {anderson1999}
\bibfield{author}{\bibinfo{person}{Edward Anderson}, \bibinfo{person}{Zhaojun
  Bai}, \bibinfo{person}{Christian Bischof}, \bibinfo{person}{Susan Blackford},
  \bibinfo{person}{Jack Dongarra}, \bibinfo{person}{Jeremy Du~Croz},
  \bibinfo{person}{Anne Greenbaum}, \bibinfo{person}{Sven Hammarling},
  \bibinfo{person}{A McKenney}, {and} \bibinfo{person}{D Sorensen}.}
  \bibinfo{year}{1999}\natexlab{}.
\newblock \bibinfo{booktitle}{{\em {LAPACK} {U}sers' guide}}.
  Vol.~\bibinfo{volume}{9}.
\newblock \bibinfo{publisher}{SIAM}.
\newblock


\bibitem[\protect\citeauthoryear{Barthels}{Barthels}{2016}]%
        {barthels2016src}
\bibfield{author}{\bibinfo{person}{Henrik Barthels}.}
  \bibinfo{year}{2016}\natexlab{}.
\newblock \showarticletitle{{A} {C}ompiler for {L}inear {A}lgebra
  {O}perations}. In \bibinfo{booktitle}{{\em SPLASH '16 Companion}}.
  \bibinfo{publisher}{ACM}, \bibinfo{address}{Amsterdam, Netherlands}.
\newblock
\showDOI{%
\url{https://doi.org/10.1145/2984043.2998539}}
\newblock
\shownote{Student Research Competition.}


\bibitem[\protect\citeauthoryear{Bezanson, Karpinski, Shah, and
  Edelman}{Bezanson et~al\mbox{.}}{2012}]%
        {bezanson2012}
\bibfield{author}{\bibinfo{person}{Jeff Bezanson}, \bibinfo{person}{Stefan
  Karpinski}, \bibinfo{person}{Viral~B. Shah}, {and} \bibinfo{person}{Alan
  Edelman}.} \bibinfo{year}{2012}\natexlab{}.
\newblock \showarticletitle{{J}ulia: {A} {F}ast {D}ynamic {L}anguage for
  {T}echnical {C}omputing}.
\newblock  (\bibinfo{date}{September} \bibinfo{year}{2012}).
\newblock
\showeprint{1209.5145}


\bibitem[\protect\citeauthoryear{Bientinesi, Gunter, and Geijn}{Bientinesi
  et~al\mbox{.}}{2008}]%
        {bientinesi2008}
\bibfield{author}{\bibinfo{person}{Paolo Bientinesi}, \bibinfo{person}{Brian
  Gunter}, {and} \bibinfo{person}{Robert A. van~de Geijn}.}
  \bibinfo{year}{2008}\natexlab{}.
\newblock \showarticletitle{{F}amilies of {A}lgorithms {R}elated to the
  {I}nversion of a {S}ymmetric {P}ositive {D}efinite {M}atrix}.
\newblock \bibinfo{journal}{{\em ACM Trans. Math. Softw.\/}}
  \bibinfo{volume}{35}, \bibinfo{number}{1}, Article \bibinfo{articleno}{3}
  (\bibinfo{date}{July} \bibinfo{year}{2008}), \bibinfo{numpages}{22}~pages.
\newblock
\showISSN{0098-3500}
\showDOI{%
\url{https://doi.org/10.1145/1377603.1377606}}


\bibitem[\protect\citeauthoryear{Bradford, Rawlins, and Shannon}{Bradford
  et~al\mbox{.}}{1998}]%
        {bradford1998}
\bibfield{author}{\bibinfo{person}{Phillip~G. Bradford},
  \bibinfo{person}{Gregory~J.E. Rawlins}, {and} \bibinfo{person}{Gregory~E.
  Shannon}.} \bibinfo{year}{1998}\natexlab{}.
\newblock \showarticletitle{{E}fficient {M}atrix {C}hain {O}rdering in
  {P}olylog {T}ime}.
\newblock \bibinfo{journal}{{\it SIAM J. Comput.}} \bibinfo{volume}{27},
  \bibinfo{number}{2} (\bibinfo{year}{1998}), \bibinfo{pages}{466--490}.
\newblock


\bibitem[\protect\citeauthoryear{Bronstein, Choukroun, Kimmel, and
  Sela}{Bronstein et~al\mbox{.}}{2016}]%
        {bronstein2016}
\bibfield{author}{\bibinfo{person}{Alexander~M. Bronstein},
  \bibinfo{person}{Yoni Choukroun}, \bibinfo{person}{Ron Kimmel}, {and}
  \bibinfo{person}{Matan Sela}.} \bibinfo{year}{2016}\natexlab{}.
\newblock \showarticletitle{{C}onsistent {D}iscretization and {M}inimization of
  the {L1} {N}orm on {M}anifolds}.
\newblock \bibinfo{journal}{{\em CoRR\/}}  \bibinfo{volume}{abs/1609.05434}
  (\bibinfo{year}{2016}).
\newblock
\showeprint{1609.05434}


\bibitem[\protect\citeauthoryear{Chin}{Chin}{1978}]%
        {Chin1978}
\bibfield{author}{\bibinfo{person}{Francis~Y. Chin}.}
  \bibinfo{year}{1978}\natexlab{}.
\newblock \showarticletitle{{A}n {O(N)} {A}lgorithm for {D}etermining a
  {N}ear-optimal {C}omputation {O}rder of {M}atrix {C}hain {P}roducts}.
\newblock \bibinfo{journal}{{\em Commun. ACM\/}} \bibinfo{volume}{21},
  \bibinfo{number}{7} (\bibinfo{date}{July} \bibinfo{year}{1978}),
  \bibinfo{pages}{544--549}.
\newblock
\showISSN{0001-0782}
\showDOI{%
\url{https://doi.org/10.1145/359545.359556}}


\bibitem[\protect\citeauthoryear{Choi, Dongarra, and Walker}{Choi
  et~al\mbox{.}}{1995}]%
        {choi1995}
\bibfield{author}{\bibinfo{person}{Jaeyoung Choi}, \bibinfo{person}{Jack~J
  Dongarra}, {and} \bibinfo{person}{David~W Walker}.}
  \bibinfo{year}{1995}\natexlab{}.
\newblock \showarticletitle{{T}he {D}esign of a {P}arallel {D}ense {L}inear
  {A}lgebra {S}oftware {L}ibrary: {R}eduction to {H}essenberg, {T}ridiagonal,
  and {B}idiagonal {F}orm}.
\newblock \bibinfo{journal}{{\em Numerical Algorithms\/}} \bibinfo{volume}{10},
  \bibinfo{number}{2} (\bibinfo{year}{1995}), \bibinfo{pages}{379--399}.
\newblock


\bibitem[\protect\citeauthoryear{Christian}{Christian}{1993}]%
        {christian1993}
\bibfield{author}{\bibinfo{person}{Jim Christian}.}
  \bibinfo{year}{1993}\natexlab{}.
\newblock \showarticletitle{{F}latterms, {D}iscrimination {N}ets, and {F}ast
  {T}erm {R}ewriting}.
\newblock \bibinfo{journal}{{\em Journal of automated reasoning\/}}
  \bibinfo{volume}{10}, \bibinfo{number}{1} (\bibinfo{year}{1993}),
  \bibinfo{pages}{95--113}.
\newblock


\bibitem[\protect\citeauthoryear{Cormen, Rivest, and Leiserson}{Cormen
  et~al\mbox{.}}{1990}]%
        {cormen1990}
\bibfield{author}{\bibinfo{person}{Thomas~H. Cormen},
  \bibinfo{person}{Ronald~L. Rivest}, {and} \bibinfo{person}{Charles~E.
  Leiserson}.} \bibinfo{year}{1990}\natexlab{}.
\newblock \bibinfo{booktitle}{{\em {I}ntroduction to {A}lgorithms}}.
\newblock \bibinfo{publisher}{McGraw-Hill, Inc.}
\newblock


\bibitem[\protect\citeauthoryear{Czumaj}{Czumaj}{1996}]%
        {czumaj1996}
\bibfield{author}{\bibinfo{person}{Artur Czumaj}.}
  \bibinfo{year}{1996}\natexlab{}.
\newblock \showarticletitle{{V}ery {F}ast {A}pproximation of the {M}atrix
  {C}hain {P}roduct {P}roblem}.
\newblock \bibinfo{journal}{{\em Journal of Algorithms\/}}
  \bibinfo{volume}{21}, \bibinfo{number}{1} (\bibinfo{year}{1996}),
  \bibinfo{pages}{71--79}.
\newblock


\bibitem[\protect\citeauthoryear{Ding and Selesnick}{Ding and
  Selesnick}{2016}]%
        {ding2016}
\bibfield{author}{\bibinfo{person}{Yin Ding} {and} \bibinfo{person}{Ivan~W.
  Selesnick}.} \bibinfo{year}{2016}\natexlab{}.
\newblock \showarticletitle{{S}parsity-{B}ased {C}orrection of {E}xponential
  {A}rtifacts}.
\newblock \bibinfo{journal}{{\em Signal Processing\/}}  \bibinfo{volume}{120}
  (\bibinfo{year}{2016}), \bibinfo{pages}{236--248}.
\newblock


\bibitem[\protect\citeauthoryear{Dongarra, Du~Croz, Hammarling, and
  Duff}{Dongarra et~al\mbox{.}}{1990}]%
        {dongarra1990}
\bibfield{author}{\bibinfo{person}{Jack~J. Dongarra}, \bibinfo{person}{Jeremy
  Du~Croz}, \bibinfo{person}{Sven Hammarling}, {and} \bibinfo{person}{Iain~S.
  Duff}.} \bibinfo{year}{1990}\natexlab{}.
\newblock \showarticletitle{{A} set of {L}evel 3 {B}asic {L}inear {A}lgebra
  {S}ubprograms}.
\newblock \bibinfo{journal}{{\em ACM Transactions on Mathematical Software
  (TOMS)\/}} \bibinfo{volume}{16}, \bibinfo{number}{1} (\bibinfo{year}{1990}),
  \bibinfo{pages}{1--17}.
\newblock


\bibitem[\protect\citeauthoryear{Dongarra, Hammarling, Hanson, and
  Croz}{Dongarra et~al\mbox{.}}{1988}]%
        {dongarra1988}
\bibfield{author}{\bibinfo{person}{Jack~J. Dongarra}, \bibinfo{person}{Sven
  Hammarling}, \bibinfo{person}{Richard~J. Hanson}, {and}
  \bibinfo{person}{Jeremy Croz}.} \bibinfo{year}{1988}\natexlab{}.
\newblock \showarticletitle{{A}n {E}xtended set of {FORTRAN} {B}asic {L}inear
  {A}lgebra {S}ubprograms}.
\newblock \bibinfo{journal}{{\em ACM Transactions on Mathematical Software
  (TOMS)\/}} \bibinfo{volume}{14}, \bibinfo{number}{4} (\bibinfo{year}{1988}),
  \bibinfo{pages}{399}.
\newblock


\bibitem[\protect\citeauthoryear{Fabregat-Traver and
  Bientinesi}{Fabregat-Traver and Bientinesi}{2011a}]%
        {fabregat-traver2011}
\bibfield{author}{\bibinfo{person}{Diego Fabregat-Traver} {and}
  \bibinfo{person}{Paolo Bientinesi}.} \bibinfo{year}{2011}\natexlab{a}.
\newblock \showarticletitle{{A}utomatic {G}eneration of {L}oop-{I}nvariants for
  {M}atrix {O}perations}. In \bibinfo{booktitle}{{\em Computational Science and
  its Applications, International Conference}}. \bibinfo{publisher}{IEEE
  Computer Society}, \bibinfo{address}{Los Alamitos, CA, USA},
  \bibinfo{pages}{82--92}.
\newblock


\bibitem[\protect\citeauthoryear{Fabregat-Traver and
  Bientinesi}{Fabregat-Traver and Bientinesi}{2011b}]%
        {FabregatTraver:2011gu}
\bibfield{author}{\bibinfo{person}{Diego Fabregat-Traver} {and}
  \bibinfo{person}{Paolo Bientinesi}.} \bibinfo{year}{2011}\natexlab{b}.
\newblock \showarticletitle{{Knowledge-Based Automatic Generation of
  Partitioned Matrix Expressions.}}
\newblock \bibinfo{journal}{{\em CASC\/}} \bibinfo{volume}{6885},
  \bibinfo{number}{4} (\bibinfo{year}{2011}), \bibinfo{pages}{144--157}.
\newblock


\bibitem[\protect\citeauthoryear{Fabregat-Traver and
  Bientinesi}{Fabregat-Traver and Bientinesi}{2013a}]%
        {fabregat-traver2013a}
\bibfield{author}{\bibinfo{person}{Diego Fabregat-Traver} {and}
  \bibinfo{person}{Paolo Bientinesi}.} \bibinfo{year}{2013}\natexlab{a}.
\newblock \showarticletitle{{A} {D}omain-{S}pecific {C}ompiler for {L}inear
  {A}lgebra {O}perations}. In \bibinfo{booktitle}{{\em High Performance
  Computing for Computational Science -- VECPAR 2010}} {\em
  (\bibinfo{series}{Lecture Notes in Computer Science})},
  \bibfield{editor}{\bibinfo{person}{O.~Marques M.~Dayde} {and}
  \bibinfo{person}{K.~Nakajima}} (Eds.), Vol.~\bibinfo{volume}{7851}.
  \bibinfo{publisher}{Springer}, \bibinfo{address}{Heidelberg},
  \bibinfo{pages}{346--361}.
\newblock


\bibitem[\protect\citeauthoryear{Fabregat-Traver and
  Bientinesi}{Fabregat-Traver and Bientinesi}{2013b}]%
        {fabregat-traver2013b}
\bibfield{author}{\bibinfo{person}{Diego Fabregat-Traver} {and}
  \bibinfo{person}{Paolo Bientinesi}.} \bibinfo{year}{2013}\natexlab{b}.
\newblock \showarticletitle{{A}pplication-tailored {L}inear {A}lgebra
  {A}lgorithms: {A} search-sased {A}pproach}.
\newblock \bibinfo{journal}{{\em International Journal of High Performance
  Computing Applications (IJHPCA)\/}} \bibinfo{volume}{27}, \bibinfo{number}{4}
  (\bibinfo{date}{Nov.} \bibinfo{year}{2013}), \bibinfo{pages}{425 -- 438}.
\newblock


\bibitem[\protect\citeauthoryear{Golub and Van~Loan}{Golub and
  Van~Loan}{2013}]%
        {golub2013}
\bibfield{author}{\bibinfo{person}{Gene~H. Golub} {and}
  \bibinfo{person}{Charles~F. Van~Loan}.} \bibinfo{year}{2013}\natexlab{}.
\newblock \bibinfo{booktitle}{{\em {M}atrix {C}omputations}}.
  Vol.~\bibinfo{volume}{4}.
\newblock \bibinfo{publisher}{Johns Hopkins}.
\newblock


\bibitem[\protect\citeauthoryear{Gr{\"a}f}{Gr{\"a}f}{1991}]%
        {graef1991}
\bibfield{author}{\bibinfo{person}{Albert Gr{\"a}f}.}
  \bibinfo{year}{1991}\natexlab{}.
\newblock \showarticletitle{{L}eft-to-{R}ight {T}ree {P}attern {M}atching}. In
  \bibinfo{booktitle}{{\em International Conference on Rewriting Techniques and
  Applications}}. Springer, \bibinfo{pages}{323--334}.
\newblock


\bibitem[\protect\citeauthoryear{Guennebaud, Jacob, et~al\mbox{.}}{Guennebaud
  et~al\mbox{.}}{2010}]%
        {eigenweb}
\bibfield{author}{\bibinfo{person}{Ga\"{e}l Guennebaud},
  \bibinfo{person}{Beno\^{i}t Jacob}, {et~al\mbox{.}}}
  \bibinfo{year}{2010}\natexlab{}.
\newblock \bibinfo{title}{{E}igen v3}.
\newblock \bibinfo{howpublished}{\url{http://eigen.tuxfamily.org}}.
  (\bibinfo{year}{2010}).
\newblock


\bibitem[\protect\citeauthoryear{Hejazi, Azimi-Abarghouyi, Makki,
  Nasiri-Kenari, and Svensson}{Hejazi et~al\mbox{.}}{2016}]%
        {hejazi2016}
\bibfield{author}{\bibinfo{person}{M. Hejazi}, \bibinfo{person}{S.~M.
  Azimi-Abarghouyi}, \bibinfo{person}{B. Makki}, \bibinfo{person}{M.
  Nasiri-Kenari}, {and} \bibinfo{person}{T. Svensson}.}
  \bibinfo{year}{2016}\natexlab{}.
\newblock \showarticletitle{{R}obust {S}uccessive {C}ompute-and-{F}orward
  {O}ver {M}ultiuser {M}ultirelay {N}etworks}.
\newblock \bibinfo{journal}{{\em IEEE Transactions on Vehicular Technology\/}}
  \bibinfo{volume}{65}, \bibinfo{number}{10} (\bibinfo{date}{Oct}
  \bibinfo{year}{2016}), \bibinfo{pages}{8112--8129}.
\newblock
\showISSN{0018-9545}
\showDOI{%
\url{https://doi.org/10.1109/TVT.2015.2506981}}


\bibitem[\protect\citeauthoryear{Hu and Shing}{Hu and Shing}{1982}]%
        {hu1982}
\bibfield{author}{\bibinfo{person}{T.C. Hu} {and} \bibinfo{person}{M.T.
  Shing}.} \bibinfo{year}{1982}\natexlab{}.
\newblock \showarticletitle{{C}omputation of {M}atrix {C}hain {P}roducts.
  {P}art {I}}.
\newblock \bibinfo{journal}{{\it SIAM J. Comput.}} \bibinfo{volume}{11},
  \bibinfo{number}{2} (\bibinfo{year}{1982}), \bibinfo{pages}{362--373}.
\newblock


\bibitem[\protect\citeauthoryear{Hu and Shing}{Hu and Shing}{1984}]%
        {hu1984}
\bibfield{author}{\bibinfo{person}{T.C. Hu} {and} \bibinfo{person}{M.T.
  Shing}.} \bibinfo{year}{1984}\natexlab{}.
\newblock \showarticletitle{{C}omputation of {M}atrix {C}hain {P}roducts.
  {P}art {II}}.
\newblock \bibinfo{journal}{{\it SIAM J. Comput.}} \bibinfo{volume}{13},
  \bibinfo{number}{2} (\bibinfo{year}{1984}), \bibinfo{pages}{228--251}.
\newblock


\bibitem[\protect\citeauthoryear{Iglberger, Hager, Treibig, and
  R{\"u}de}{Iglberger et~al\mbox{.}}{2012}]%
        {iglberger2012a}
\bibfield{author}{\bibinfo{person}{Klaus Iglberger}, \bibinfo{person}{Georg
  Hager}, \bibinfo{person}{Jan Treibig}, {and} \bibinfo{person}{Ulrich
  R{\"u}de}.} \bibinfo{year}{2012}\natexlab{}.
\newblock \showarticletitle{{E}xpression {T}emplates {R}evisited: {A}
  {P}erformance {A}nalysis of the {C}urrent {ET} {M}ethodologies}.
\newblock \bibinfo{journal}{{\em SIAM Journal on Scientific Computing\/}}
  \bibinfo{volume}{34}, \bibinfo{number}{2} (\bibinfo{year}{2012}),
  \bibinfo{pages}{C42--C69}.
\newblock


\bibitem[\protect\citeauthoryear{Krebber}{Krebber}{2017}]%
        {krebber2017}
\bibfield{author}{\bibinfo{person}{Manuel Krebber}.}
  \bibinfo{year}{2017}\natexlab{}.
\newblock \showarticletitle{{N}on-linear {A}ssociative-{C}ommutative
  {M}any-to-{O}ne {P}attern {M}atching with {S}equence {V}ariables}.
\newblock \bibinfo{journal}{{\em CoRR\/}}  \bibinfo{volume}{abs/1705.00907}
  (\bibinfo{year}{2017}).
\newblock
\showURL{%
\url{http://arxiv.org/abs/1705.00907}}


\bibitem[\protect\citeauthoryear{Lawson, Hanson, Kincaid, and Krogh}{Lawson
  et~al\mbox{.}}{1979}]%
        {lawson1979}
\bibfield{author}{\bibinfo{person}{Chuck~L. Lawson},
  \bibinfo{person}{Richard~J. Hanson}, \bibinfo{person}{David~R. Kincaid},
  {and} \bibinfo{person}{Fred~T. Krogh}.} \bibinfo{year}{1979}\natexlab{}.
\newblock \showarticletitle{{B}asic {L}inear {A}lgebra {S}ubprograms for
  {FORTRAN} {U}sage}.
\newblock \bibinfo{journal}{{\em ACM Transactions on Mathematical Software
  (TOMS)\/}} \bibinfo{volume}{5}, \bibinfo{number}{3} (\bibinfo{year}{1979}),
  \bibinfo{pages}{308--323}.
\newblock


\bibitem[\protect\citeauthoryear{Lee, Kim, Hong, and Lee}{Lee
  et~al\mbox{.}}{2003}]%
        {lee2003}
\bibfield{author}{\bibinfo{person}{Heejo Lee}, \bibinfo{person}{Jong Kim},
  \bibinfo{person}{Sung~Je Hong}, {and} \bibinfo{person}{Sunggu Lee}.}
  \bibinfo{year}{2003}\natexlab{}.
\newblock \showarticletitle{{P}rocessor {A}llocation and {T}ask {S}cheduling of
  {M}atrix {C}hain {P}roducts on {P}arallel {S}ystems}.
\newblock \bibinfo{journal}{{\em Parallel and Distributed Systems, IEEE
  Transactions on\/}} \bibinfo{volume}{14}, \bibinfo{number}{4}
  (\bibinfo{year}{2003}), \bibinfo{pages}{394--407}.
\newblock


\bibitem[\protect\citeauthoryear{Nachtigal, Reddy, and Trefethen}{Nachtigal
  et~al\mbox{.}}{1992}]%
        {nachtigal1992}
\bibfield{author}{\bibinfo{person}{No{\"e}l~M Nachtigal},
  \bibinfo{person}{Satish~C Reddy}, {and} \bibinfo{person}{Lloyd~N Trefethen}.}
  \bibinfo{year}{1992}\natexlab{}.
\newblock \showarticletitle{{H}ow {F}ast are {N}onsymmetric {M}atrix
  {I}terations?}
\newblock \bibinfo{journal}{{\it SIAM J. Matrix Anal. Appl.}}
  \bibinfo{volume}{13}, \bibinfo{number}{3} (\bibinfo{year}{1992}),
  \bibinfo{pages}{778--795}.
\newblock


\bibitem[\protect\citeauthoryear{Nedjah, Walter, and Eldridge}{Nedjah
  et~al\mbox{.}}{1997}]%
        {nedjah1997}
\bibfield{author}{\bibinfo{person}{Nadia Nedjah}, \bibinfo{person}{Colin
  Walter}, {and} \bibinfo{person}{Stephen Eldridge}.}
  \bibinfo{year}{1997}\natexlab{}.
\newblock \showarticletitle{Optimal {L}eft-to-{R}ight {P}attern-{M}atching
  {A}utomata}. In \bibinfo{booktitle}{{\em Algebraic and Logic Programming}}.
  Springer, \bibinfo{pages}{273--286}.
\newblock


\bibitem[\protect\citeauthoryear{Ni{\~{n}}o, Sandu, and Deng}{Ni{\~{n}}o
  et~al\mbox{.}}{2016}]%
        {nino2016}
\bibfield{author}{\bibinfo{person}{Elias~D. Ni{\~{n}}o},
  \bibinfo{person}{Adrian Sandu}, {and} \bibinfo{person}{Xinwei Deng}.}
  \bibinfo{year}{2016}\natexlab{}.
\newblock \showarticletitle{{A} {P}arallel {I}mplementation of the {E}nsemble
  {K}alman {F}ilter {B}ased on {M}odified {C}holesky {D}ecomposition}.
\newblock \bibinfo{journal}{{\em CoRR\/}}  \bibinfo{volume}{abs/1606.00807}
  (\bibinfo{year}{2016}).
\newblock
\showURL{%
\url{http://arxiv.org/abs/1606.00807}}


\bibitem[\protect\citeauthoryear{Nishida, Ito, and Nakano}{Nishida
  et~al\mbox{.}}{2011}]%
        {nishida2011}
\bibfield{author}{\bibinfo{person}{Kazufumi Nishida}, \bibinfo{person}{Yasuaki
  Ito}, {and} \bibinfo{person}{Koji Nakano}.} \bibinfo{year}{2011}\natexlab{}.
\newblock \showarticletitle{{A}ccelerating the {D}ynamic {P}rogramming for the
  {M}atrix {C}hain {P}roduct on the {GPU}}. In \bibinfo{booktitle}{{\em
  Networking and Computing (ICNC), 2011 Second International Conference on}}.
  IEEE, \bibinfo{pages}{320--326}.
\newblock


\bibitem[\protect\citeauthoryear{Noschese and Reichel}{Noschese and
  Reichel}{2016}]%
        {noschese2016}
\bibfield{author}{\bibinfo{person}{Silvia Noschese} {and}
  \bibinfo{person}{Lothar Reichel}.} \bibinfo{year}{2016}\natexlab{}.
\newblock \showarticletitle{{S}ome {M}atrix {N}earness {P}roblems {S}uggested
  by {T}ikhonov {R}egularization}.
\newblock \bibinfo{journal}{{\it Linear Algebra Appl.}}  \bibinfo{volume}{502}
  (\bibinfo{year}{2016}), \bibinfo{pages}{366--386}.
\newblock


\bibitem[\protect\citeauthoryear{Parikh, Myers, and van~de Geijn}{Parikh
  et~al\mbox{.}}{2017}]%
        {parikh2017}
\bibfield{author}{\bibinfo{person}{Devangi~N Parikh}, \bibinfo{person}{Maggie~E
  Myers}, {and} \bibinfo{person}{Robert~A van~de Geijn}.}
  \bibinfo{year}{2017}\natexlab{}.
\newblock \showarticletitle{Deriving Correct High-Performance Algorithms}.
\newblock \bibinfo{journal}{{\em arXiv preprint arXiv:1710.04286\/}}
  (\bibinfo{year}{2017}).
\newblock
\showeprint{1710.04286}


\bibitem[\protect\citeauthoryear{Peise and Bientinesi}{Peise and
  Bientinesi}{2012}]%
        {peise2012}
\bibfield{author}{\bibinfo{person}{Elmar Peise} {and} \bibinfo{person}{Paolo
  Bientinesi}.} \bibinfo{year}{2012}\natexlab{}.
\newblock \showarticletitle{{P}erformance {M}odeling for {D}ense {L}inear
  {A}lgebra}. In \bibinfo{booktitle}{{\em Proceedings of the 2012 SC Companion:
  High Performance Computing, Networking Storage and Analysis (PMBS12)}} {\em
  (\bibinfo{series}{SCC '12})}. \bibinfo{publisher}{IEEE Computer Society},
  \bibinfo{address}{Washington, DC, USA}, \bibinfo{pages}{406--416}.
\newblock


\bibitem[\protect\citeauthoryear{Peise and Bientinesi}{Peise and
  Bientinesi}{2015a}]%
        {peise2015a}
\bibfield{author}{\bibinfo{person}{Elmar Peise} {and} \bibinfo{person}{Paolo
  Bientinesi}.} \bibinfo{year}{2015}\natexlab{a}.
\newblock \showarticletitle{{A} {S}tudy on the {I}nfluence of {C}aching:
  {S}equences of {D}ense {L}inear {A}lgebra {K}ernels}. In
  \bibinfo{booktitle}{{\em High Performance Computing for Computational Science
  -- VECPAR 2014}} {\em (\bibinfo{series}{Lecture Notes in Computer Science})},
  \bibfield{editor}{\bibinfo{person}{Michel Dayd\'{e}}, \bibinfo{person}{Osni
  Marques}, {and} \bibinfo{person}{Kengo Nakajima}} (Eds.),
  Vol.~\bibinfo{volume}{8969}. \bibinfo{publisher}{Springer International
  Publishing}, \bibinfo{pages}{245--258}.
\newblock
\showeprint{1402.5897v1}


\bibitem[\protect\citeauthoryear{Peise and Bientinesi}{Peise and
  Bientinesi}{2015b}]%
        {peise2015}
\bibfield{author}{\bibinfo{person}{Elmar Peise} {and} \bibinfo{person}{Paolo
  Bientinesi}.} \bibinfo{year}{2015}\natexlab{b}.
\newblock \showarticletitle{{T}he {ELAPS} {F}ramework: {E}xperimental {L}inear
  {A}lgebra {P}erformance {S}tudies}.
\newblock \bibinfo{journal}{{\em CoRR\/}}  \bibinfo{volume}{abs/1504.08035}
  (\bibinfo{year}{2015}).
\newblock
\showURL{%
\url{http://arxiv.org/abs/1504.08035}}


\bibitem[\protect\citeauthoryear{Ramanan}{Ramanan}{1996}]%
        {ramanan1996}
\bibfield{author}{\bibinfo{person}{Prakash Ramanan}.}
  \bibinfo{year}{1996}\natexlab{}.
\newblock \showarticletitle{{A}n {E}fficient {P}arallel {A}lgorithm for the
  {M}atrix-{C}hain-{P}roduct {P}roblem}.
\newblock \bibinfo{journal}{{\it SIAM J. Comput.}} \bibinfo{volume}{25},
  \bibinfo{number}{4} (\bibinfo{year}{1996}), \bibinfo{pages}{874--893}.
\newblock


\bibitem[\protect\citeauthoryear{{Rao}, {Sandu}, {Ng}, and {Nino-Ruiz}}{{Rao}
  et~al\mbox{.}}{2015}]%
        {rao2015}
\bibfield{author}{\bibinfo{person}{V. {Rao}}, \bibinfo{person}{A. {Sandu}},
  \bibinfo{person}{M. {Ng}}, {and} \bibinfo{person}{E. {Nino-Ruiz}}.}
  \bibinfo{year}{2015}\natexlab{}.
\newblock \showarticletitle{{{R}obust {D}ata {A}ssimilation {U}sing $L\_1$ and
  Huber {N}orms}}.
\newblock \bibinfo{journal}{{\em ArXiv e-prints\/}} (\bibinfo{date}{Nov.}
  \bibinfo{year}{2015}).
\newblock
\showeprint[arxiv]{math.NA/1511.01593}


\bibitem[\protect\citeauthoryear{Rao, Sandu, Ng, and Nino-Ruiz}{Rao
  et~al\mbox{.}}{2017}]%
        {rao2015:siam}
\bibfield{author}{\bibinfo{person}{Vishwas Rao}, \bibinfo{person}{Adrian
  Sandu}, \bibinfo{person}{Michael Ng}, {and} \bibinfo{person}{Elias~D.
  Nino-Ruiz}.} \bibinfo{year}{2017}\natexlab{}.
\newblock \showarticletitle{{R}obust {D}ata {A}ssimilation {U}sing $L_1$ and
  Huber {N}orms}.
\newblock \bibinfo{journal}{{\em SIAM Journal on Scientific Computing\/}}
  \bibinfo{volume}{39}, \bibinfo{number}{3} (\bibinfo{year}{2017}),
  \bibinfo{pages}{B548--B570}.
\newblock
\showDOI{%
\url{https://doi.org/10.1137/15M1045910}}


\bibitem[\protect\citeauthoryear{Ronellenfitsch, Timme, and
  Witthaut}{Ronellenfitsch et~al\mbox{.}}{2015}]%
        {ronellenfitsch2015}
\bibfield{author}{\bibinfo{person}{Henrik Ronellenfitsch},
  \bibinfo{person}{Marc Timme}, {and} \bibinfo{person}{Dirk Witthaut}.}
  \bibinfo{year}{2015}\natexlab{}.
\newblock \showarticletitle{{A} {D}ual {M}ethod for {C}omputing {P}ower
  {T}ransfer {D}istribution {F}actors}.
\newblock \bibinfo{journal}{{\em CoRR\/}}  \bibinfo{volume}{abs/1510.04645}
  (\bibinfo{year}{2015}).
\newblock
\showURL{%
\url{http://arxiv.org/abs/1510.04645}}


\bibitem[\protect\citeauthoryear{Sanderson}{Sanderson}{2010}]%
        {sanderson2010}
\bibfield{author}{\bibinfo{person}{Conrad Sanderson}.}
  \bibinfo{year}{2010}\natexlab{}.
\newblock \showarticletitle{{A}rmadillo: {A}n {O}pen {S}ource {C}++ {L}inear
  {A}lgebra {L}ibrary for {F}ast {P}rototyping and {C}omputationally
  {I}ntensive {E}xperiments}.
\newblock  (\bibinfo{year}{2010}).
\newblock


\bibitem[\protect\citeauthoryear{Siek, Karlin, and Jessup}{Siek
  et~al\mbox{.}}{2008}]%
        {siek2008}
\bibfield{author}{\bibinfo{person}{Jeremy~G. Siek}, \bibinfo{person}{Ian
  Karlin}, {and} \bibinfo{person}{Elizabeth~R. Jessup}.}
  \bibinfo{year}{2008}\natexlab{}.
\newblock \showarticletitle{{B}uild to {O}rder {L}inear {A}lgebra {K}ernels}.
  In \bibinfo{booktitle}{{\em Parallel and Distributed Processing, 2008. IPDPS
  2008. IEEE International Symposium on}}. IEEE, \bibinfo{pages}{1--8}.
\newblock


\bibitem[\protect\citeauthoryear{Spampinato and P{\"u}schel}{Spampinato and
  P{\"u}schel}{2014}]%
        {spampinato2014}
\bibfield{author}{\bibinfo{person}{Daniele~G. Spampinato} {and}
  \bibinfo{person}{Markus P{\"u}schel}.} \bibinfo{year}{2014}\natexlab{}.
\newblock \showarticletitle{{A} {B}asic {L}inear {A}lgebra {C}ompiler}. In
  \bibinfo{booktitle}{{\em Proceedings of Annual IEEE/ACM International
  Symposium on Code Generation and Optimization}}. ACM, \bibinfo{pages}{23}.
\newblock


\bibitem[\protect\citeauthoryear{Spampinato and P{\"u}schel}{Spampinato and
  P{\"u}schel}{2016}]%
        {spampinato2016}
\bibfield{author}{\bibinfo{person}{Daniele~G Spampinato} {and}
  \bibinfo{person}{Markus P{\"u}schel}.} \bibinfo{year}{2016}\natexlab{}.
\newblock \showarticletitle{{A} {B}asic {L}inear {A}lgebra {C}ompiler for
  {S}tructured {M}atrices}. In \bibinfo{booktitle}{{\em International Symposium
  on Code Generation and Optimization (CGO)}}. \bibinfo{pages}{117--127}.
\newblock


\bibitem[\protect\citeauthoryear{Straszak and Vishnoi}{Straszak and
  Vishnoi}{2015}]%
        {straszak2015}
\bibfield{author}{\bibinfo{person}{Damian Straszak} {and}
  \bibinfo{person}{Nisheeth~K Vishnoi}.} \bibinfo{year}{2015}\natexlab{}.
\newblock \showarticletitle{{O}n a {N}atural {D}ynamics for {L}inear
  {P}rogramming}.
\newblock  (\bibinfo{year}{2015}).
\newblock
\showeprint{1511.07020}


\bibitem[\protect\citeauthoryear{Strate, Wainwright, et~al\mbox{.}}{Strate
  et~al\mbox{.}}{1990}]%
        {strate1990}
\bibfield{author}{\bibinfo{person}{Steve Strate}, \bibinfo{person}{Roger~L.
  Wainwright}, {et~al\mbox{.}}} \bibinfo{year}{1990}\natexlab{}.
\newblock \showarticletitle{{P}arallelization of the {D}ynamic {P}rogramming
  {A}lgorithm for the {M}atrix {C}hain {P}roduct on a {H}ypercube}. In
  \bibinfo{booktitle}{{\em Applied Computing, 1990., Proceedings of the 1990
  Symposium on}}. IEEE, \bibinfo{pages}{78--84}.
\newblock


\bibitem[\protect\citeauthoryear{Veldhuizen}{Veldhuizen}{1998}]%
        {veldhuizen1998}
\bibfield{author}{\bibinfo{person}{Todd~L. Veldhuizen}.}
  \bibinfo{year}{1998}\natexlab{}.
\newblock \showarticletitle{{A}rrays in {B}litz++}. In \bibinfo{booktitle}{{\em
  International Symposium on Computing in Object-Oriented Parallel
  Environments}}. Springer, \bibinfo{pages}{223--230}.
\newblock


\end{thebibliography}
